%% file: neurips_2026.tex
\documentclass{article}

\usepackage[main, final]{neurips_2026}

\usepackage[utf8]{inputenc} 
\usepackage[T1]{fontenc}    
\usepackage{hyperref}       
\usepackage{url}            
\usepackage{booktabs}       
\usepackage{amsfonts}       
\usepackage{nicefrac}       
\usepackage{microtype}      
\usepackage{xcolor}         
\usepackage{amssymb}
\usepackage{pifont}
\usepackage{graphicx} 
\usepackage{multirow}
\usepackage{makecell}
\usepackage{enumitem}
\usepackage{tcolorbox}
\usepackage{amsmath}
\tcbuselibrary{breakable}
\tcbuselibrary{listings}

\definecolor{figred}{RGB}{243,161,158}
\definecolor{figgreen}{RGB}{0,128,0}

\newcommand{\method}{{\textsc{SAFTAC}}}

\newcommand{\xmark}{\ding{55}}
\title{SAFTAC: Simulation-Augmented Fine-Tuning of Open-Source LLMs for Analog Circuit Design}

\author{%
  Junsheng Huang\thanks{Equal contribution. Correspondence to Junsheng Huang <\texttt{jh103@illinois.edu}>, Bin Hu <\texttt{binhu7@illinois.edu}>, and Huan Zhang <\texttt{huan@huan-zhang.com}>.}\quad~Yifan Sun\footnotemark[1]\quad~Zhuoer Zhang\footnotemark[1]\quad~Ning Wei\quad~Yening Liu\quad~Michael Molter
  \\
  \textbf{Pavan Kumar Hanumolu}\quad~\textbf{Elyse Rosenbaum}\quad~\textbf{Bin Hu}\quad~\textbf{Huan Zhang}\\
  University of Illinois Urbana-Champaign \\ 
}

\begin{document}

\maketitle

\input{Sec/0_abstract}
\input{Sec/1_introduction}
\input{Sec/2_preliminary_related_work}

\input{Sec/3_dataset}

\input{Sec/4_method}
\input{Sec/5_experiment}

\input{Sec/6_conclusion}

\bibliographystyle{plain}
\small\bibliography{neurips_2026}


\input{Sec/7_appendix}


\newpage

\input{Sec/8_checklist}
\end{document}

%% file: Sec/0_abstract.tex
\begin{abstract}
Circuit design for analog integrated circuits (ICs) is challenging because circuit topology, device sizing, and circuit performance are tightly coupled. Existing large language model (LLM)-based methods either rely on inference-time correction around frozen models or train on limited circuit types without directly using simulation outcomes as training signals. A key bottleneck is that existing datasets do not provide sufficient task information for training LLMs on end-to-end specification-conditioned design. To fill this gap, we first construct a large-scale simulation-grounded dataset with 8,626 design tasks across 10 different types of analog circuits, where each task includes a textual description, labeled I/O ports, loading condition, target specifications, a testbench, and a reference netlist. Using this dataset, we present \method{}, a simulation-augmented fine-tuning framework that trains LLMs to generate and revise analog circuit design. \method{} combines three-stage supervised fine-tuning, which progressively builds the capabilities required for analog circuit design and feedback-guided correction, with simulation feedback-augmented reinforcement fine-tuning, which further optimizes models using simulation-grounded rewards. We evaluate \method{} under single-pass generation and two-pass generation with simulation feedback, showing improvements over the strongest closed-source LLM by 15.9\% and 7.8\%, respectively.
\end{abstract}

%% file: Sec/1_introduction.tex
\section{Introduction}

Analog integrated circuits (IC), such as amplifiers, oscillators, and voltage regulators, are used to process electrical signals that are continuous in both time and amplitude. A typical process of circuit design for analog ICs first defines the circuit topological structure, then selects concrete device and component values, and finally uses SPICE~\cite{nagel1973spice} simulation to verify whether the circuit satisfies the performance specifications. This is a fundamentally difficult task because circuit topology, component value assignments, and circuit performance are tightly coupled, and analog circuit design must navigate trade-offs among often competing specifications. This contrasts with circuit design for digital ICs~\cite{blocklove2023chip, chang2023chipgpt,liu2023chipnemo}, where voltages can often be abstracted as binary values and correctness can be checked through discrete functional equivalence. Consequently, off-the-shelf general-purpose large language models (LLMs) struggle to reliably satisfy stringent specifications in analog circuit design tasks, as they are not trained to reason jointly over circuit topology design, device and component value assignment, and simulation-grounded performance constraints. 


One line of recent work adapts LLMs to analog circuit design through inference-time correction, external memory, or predefined topology libraries. These training-free methods~\cite{lai2025analogcoder, lai2026analogcoder, zhang2025analogxpert, wang2025analogsage} keep the underlying LLM frozen: they can improve the surrounding design workflow, but do not turn design experience or simulation feedback into improved model capability. Moreover, in industrial settings, reliance on external APIs can limit autonomy and flexibility, since proprietary circuit data, testbenches, and simulation pipelines may need to be exposed outside the company. Training-based methods~\cite{gao2025analoggenie, chang2024lamagic, chang2025lamagic2, chen2024artisan}, on the other hand, adapt LLMs for analog circuit design, but many focus on unconditional topology synthesis rather than specification-conditioned analog circuit design. Others are restricted to limited circuit types or rely on simplified reward signals without directly utilizing SPICE simulation outcomes. For example, the ICML 2025 work AutoCircuit-RL~\cite{vijayaraghavan2025autocircuit} is closely related to ours, as it also uses instruction tuning and reinforcement learning (RL). However, it focuses on a single circuit family, i.e., switching power converters, and guides RL with learned reward models rather than direct SPICE simulation outcomes.

A fundamental bottleneck of training-based methods for automated analog circuit design is the lack of large-scale, high-quality design-task datasets. Ideally, such a dataset should cover diverse circuit types, contain sufficient samples within each type, and provide complete task components, including textual description, target specifications, labeled I/O ports, a testbench, and circuit load. Existing datasets~\cite{mehradfar2024aicircuit, li2024analoggym, bhandari2024masala} are mostly limited in size, narrow in circuit coverage, or incomplete in task information, making them insufficient for training LLMs to generate and revise analog circuit designs under simulation-based feedback or simulation-grounded rewards.


To address these challenges, we develop a simulation-grounded data curation and LLM fine-tuning pipeline for analog circuit design. We first construct a large-scale analog circuit design-task dataset through an automated data curation pipeline that transforms raw circuit sources into complete design tasks. Each task includes a textual description, target specifications, labeled I/O ports, a testbench, circuit load, and a reference netlist. The resulting dataset contains \textbf{8,626} design tasks across \textbf{10} common analog circuit types. Building on this dataset, we propose \textbf{\method{}} (\textbf{S}imulation-\textbf{A}ugmented \textbf{F}ine-\textbf{T}uning for \textbf{A}nalog \textbf{C}ircuit design), which combines three-stage supervised fine-tuning (SFT) with simulation feedback-augmented reinforcement fine-tuning (RFT). The SFT stages progressively teach circuit-domain knowledge, component-level understanding, and simulation feedback-guided correction. The RFT stage further optimizes the model using rewards from a full circuit validation pipeline, including circuit topology check, Bayesian Optimization (BO)-based component sizing, and SPICE simulation. In addition to the standard netlist generation task, we include a simulation feedback-guided correction task, where the model revises failed netlists using rich diagnostic feedback from the validation pipeline. This enables the model to generate initial circuit designs and repair failed designs toward functional, specification-satisfying analog circuits.

\textbf{In summary, our work makes the following contributions:}
\begin{itemize}[wide, labelwidth=!, noitemsep, labelindent=0pt,topsep=-\parskip]
    \item We construct a large-scale, high-quality, simulation-grounded dataset for analog circuit design. The dataset contains \textbf{8,626} design tasks across \textbf{10} common analog circuit types in analog ICs, with complete task information and human expert verification.

    \item We propose \method{}, a simulation-augmented fine-tuning pipeline that integrates three-stage SFT with simulation feedback-augmented RFT, enabling LLMs to learn both netlist generation and simulation feedback-guided correction.

    \item We apply \method{} to Qwen3-8B on $5$ representative analog circuit types, and evaluate it under two practical design settings. Experiments show that \method{} substantially improves the overall success rate, outperforms strong closed-source LLMs by 15.9\% and 7.8\% in two evaluation settings respectively, generalizes across different backbone models, remains compatible with training-free methods, and shows promising out-of-distribution performance on held-out circuit types.
\end{itemize}

%% file: Sec/2_preliminary_related_work.tex
\section{Preliminary and Related Work}

\subsection{Preliminaries on Analog Circuit Design Tasks}

\paragraph{Task Inputs.}
A circuit design task for analog ICs specifies what circuit should be built and how it will be evaluated. 
The input consists of four components: 
(1) a natural-language description of the desired circuit functionality, 
(2) labeled I/O ports for input signals, output signals and supply connections,
(3) circuit load such as capacitive or resistive output load, and 
(4) target performance specifications such as gain, bandwidth, phase margin, and power consumption.

\paragraph{Topology Generation, Sizing, and Simulation.}
Given these task inputs, the model is asked to generate a SPICE-compatible netlist that describes the circuit as interconnected components such as transistors, resistors, and capacitors. 
The generated netlist defines the discrete topology, including component types and node connections, while sizing determines the continuous parameters of that fixed topology, such as transistor dimensions, resistor values, and capacitor values.
Since analog circuit performance depends strongly on these continuous sizing parameters, a generated topology cannot be evaluated by structure alone. 
We therefore fix the topology and optimize its sizing parameters using Bayesian Optimization. 
Each candidate sizing configuration is evaluated by SPICE simulation under the task testbench and circuit load to obtain the specifications used for validation.

\paragraph{Design Success Criterion.}
A generated design is considered successful only if it passes the full simulation-based validation pipeline. 
First, the netlist must pass a circuit topology check that enforces SPICE syntax, port consistency, and topology-level
connectivity. 
Second, it must run successfully in the provided testbench. 
Third, after sizing, its simulated performance must satisfy all target specifications. 
When a design fails, the validation pipeline also provides \textit{rich} diagnostic feedback, such as syntax errors, simulation failures, or unmet specifications. This feedback is more informative than a binary success indicator and naturally supports subsequent correction.

\subsection{Related Work}
We review prior datasets and LLM-based methods for analog circuit design. Existing LLM-based design methods can be broadly categorized into training-free and training-based approaches.

\paragraph{Datasets for Analog Circuit Design.}
Existing datasets provide useful resources for analog circuit design, but they are not yet sufficient for training LLMs on end-to-end specification-conditioned design tasks. Some datasets are limited in scale, such as AnalogCoder-Pro~\cite{lai2026analogcoder}, whose evaluation set covers only 28 circuits, and AICircuit~\cite{mehradfar2024aicircuit}, which covers 7 basic circuits and 2 wireless transceiver systems. Others cover only a narrow range of circuit types, including AnalogGym~\cite{li2024analoggym}, which contains only 5 circuit types, and OCB~\cite{dong2023cktgnn}, which focuses on operational amplifiers. Larger-scale datasets such as Masala-CHAI~\cite{bhandari2024masala} and AnalogGenie~\cite{gao2025analoggenie} contain thousands of circuit examples, but mainly provide circuit images, netlists, or brief textual descriptions. They generally lack the target specifications and testbenches required for end-to-end analog circuit design training.

\paragraph{Training-Free Analog Circuit Design Methods.}
    Training-free LLM methods apply off-the-shelf LLMs to analog circuit design through structured design workflows, external design knowledge, and inference-time feedback. AnalogCoder~\cite{lai2025analogcoder} uses structured agentic workflows with simulation-based feedback loops to iteratively generate, diagnose, and correct circuits, with AnalogCoder-Pro~\cite{lai2026analogcoder} further incorporating waveform-based multimodal diagnosis. AnalogXpert~\cite{zhang2025analogxpert} uses a predefined subcircuit library to reduce the design space and decomposes topology synthesis into block selection and connection graph construction. AnalogSAGE~\cite{wang2025analogsage} develops a self-evolving multi-agent framework that uses stratified memory and simulation-grounded feedback for iterative refinement. While these methods demonstrate the potential of general-purpose LLMs for analog circuit design, they are fundamentally constrained by the frozen underlying model. They can improve the inference workflow, but cannot turn design experience and simulation feedback into improved model capability. 


\paragraph{Training-Based Analog Circuit Design Methods.}
Training-based LLM methods aim to improve analog circuit generation by training models with domain-specific data. Practically, training-based methods applied to open-source LLMs offer greater flexibility and autonomy for industrial analog circuit design. They can be adapted within a company's own design flow, reducing reliance on external APIs and limiting exposure of proprietary circuit data, testbenches, and simulation pipelines. AnalogGenie~\cite{gao2025analoggenie} constructs an analog circuit topology dataset and introduces a sequence-based graph representation for topology generation. LaMAGIC~\cite{chang2024lamagic} and LaMAGIC2~\cite{chang2025lamagic2} train language models with structured circuit formulations for power converter topology generation. Artisan~\cite{chen2024artisan} is a domain-specific LLM framework for automated operational amplifier design. AutoCircuit-RL~\cite{vijayaraghavan2025autocircuit} combines instruction tuning and RL for switching power converter topology generation, using learned reward models rather than direct simulation-derived rewards. 

Despite these efforts, existing training-based methods remain limited in scope. Some focus on unconditional topology generation rather than specification-conditioned analog circuit design, while others are restricted to specific circuit families or small topology spaces. Moreover, existing pipelines often lack a direct mechanism for using SPICE~\cite{nagel1973spice} simulation outcomes as training signals, which are crucial for optimizing models toward functional circuits that meet target specifications. Given these gaps, we develop a complete simulation-guided fine-tuning pipeline for adapting open-source LLMs to analog circuit design. We first curate a large-scale and comprehensive analog circuit design dataset with target specifications, testbenches, and circuit load. We then use this dataset to support SFT and simulation feedback-augmented RFT for circuit design and feedback-guided correction.


%% file: Sec/3_dataset.tex
\section{Dataset}

\begin{figure}[t]
    \centering
    \resizebox{\textwidth}{!}{\includegraphics{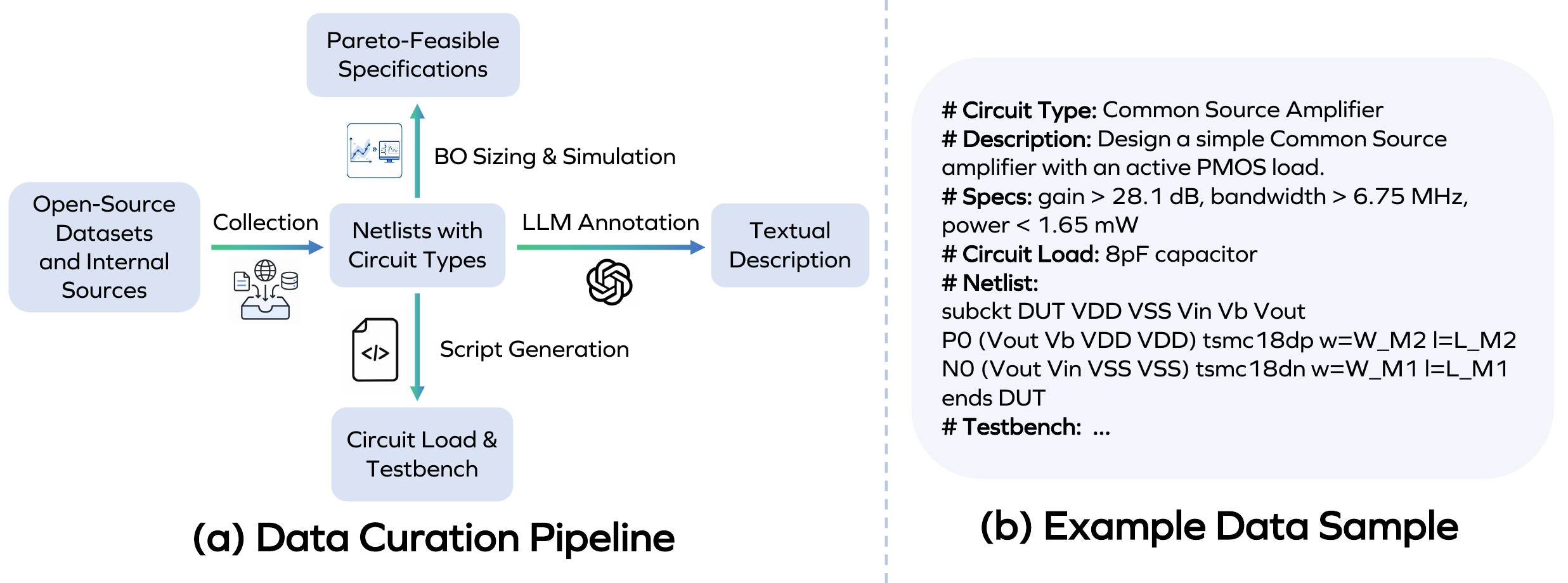}}
    \caption{\textbf{Overview of the automated data generation pipeline for analog circuit design tasks.}
(a) Raw circuit sources are collected and transformed into netlists with circuit-type labels, then augmented with textual descriptions, circuit load, testbenches, and Pareto-feasible specifications derived through BO sizing and SPICE simulation. Each step is guided and checked by human experts.
(b) A data sample is provided for reference, with the Python testbench code omitted for brevity.}
    \label{fig:data_pipeline}
\end{figure}

\begin{table}[t]
\centering
\small
\caption{\textbf{Comparison with existing analog circuit design datasets.}
We compare both task components and dataset scale. Our dataset provides the most complete set of task components while covering 10 circuit types and 8,626 design tasks. Masala-CHAI~\cite{bhandari2024masala} does not disclose circuit-type information, so its number of circuit types and average circuits per type are marked as unavailable.}
\label{tab:dataset_comparison}
\resizebox{\linewidth}{!}{
\begin{tabular}{lcccccc}
\toprule
\textbf{Feature} & \textbf{Analogcoder-Pro~\cite{lai2026analogcoder}} & \textbf{AnalogGym~\cite{li2024analoggym}} & \textbf{AMSNet-KG~\cite{shi2025amsnet}} & \textbf{AnalogGenie~\cite{gao2025analoggenie}} & \textbf{Masala-CHAI~\cite{bhandari2024masala}} & \textbf{Ours} \\
\midrule
Text Descriptions & \checkmark & \xmark & \checkmark & \xmark & \checkmark & \checkmark \\
Multiple Specs & \checkmark & \checkmark & \checkmark & \xmark & \xmark & \checkmark \\
Labeled I/O & \checkmark & \checkmark & \checkmark & \xmark & \xmark & \checkmark \\
Testbench & \checkmark & \checkmark & \checkmark & \xmark & \xmark & \checkmark \\
Circuit Load & \xmark & \checkmark & \checkmark & \xmark & \xmark & \checkmark \\
Open-Source & \checkmark & \checkmark & \xmark & \checkmark & \checkmark & \checkmark \\
\midrule
\# Circuit Types & \textbf{15} & 5 & 5 & 11 & -- & 10 \\
\# Circuits Samples & 28 & 30 & 894 & 3,350 & 7,500 & \textbf{8,626} \\
Avg. Samples/Type & 1.9 & 6.0 & 178.8 & 304.5 & -- & \textbf{862.6} \\
\bottomrule
\end{tabular}}
\end{table}

\subsection{Automated Generation of Simulation-Grounded Analog Circuit Design Tasks}

High-quality analog circuit design datasets are difficult to construct at scale. Circuit examples from open-source databases typically do not come with circuit-type annotations, testbenches, or appropriate target specifications, yet these elements are necessary for turning raw circuits into usable training tasks. Manually collecting them requires substantial expert effort and does not scale well across diverse circuit types. Conversely, naively prompting LLMs to generate such data can produce invalid circuits, non-functional testbenches, or unrealistic specifications. This is because simulation under appropriate sizing, circuit load and design constraints is an integral part of analog circuit design.

To address this challenge, we develop an automated simulation-grounded data construction pipeline, shown in Fig.~\ref{fig:data_pipeline}, that balances scalability and data quality through simulation-based validation and human circuit-expert knowledge. The pipeline begins by collecting circuit examples from the open-source datasets and internal sources. The collected circuits are manually verified for schematic–netlist consistency, connectivity, I/O ports, and circuit-type labels. Starting from these validated netlists, it constructs well-specified design tasks by (1) using an LLM annotator to produce textual design descriptions, (2) assigning circuit loads and instantiating testbench templates using deterministic Python scripts, and (3) performing BO sizing under SPICE simulation to derive Pareto-feasible target specification tuples. Circuit experts further review the LLM-generated descriptions and validate the testbench to ensure data quality. Detailed testbench implementations are
provided in App.~\ref{app:testbench}.

\subsection{Dataset Statistics}

The resulting dataset contains \textbf{8,626} tasks covering \textbf{10} circuit types, derived from \textbf{114} unique netlists. Tab.~\ref{tab:dataset_statistics} in App.~\ref{app:circuit_type} summarizes the dataset statistics by circuit type, including the performance specifications and number of samples. These statistics show that our dataset provides broad coverage across analog circuit types, with many samples in each type covering different target specifications. 

Tab.~\ref{tab:dataset_comparison} compares our dataset with prior analog circuit datasets. Beyond its coverage and scale, our dataset provides the most comprehensive task information: each sample includes a textual description, target specifications, labeled I/O ports, a testbench, circuit load, and a reference netlist. These components make the dataset directly usable for both simulation-driven SFT and RFT, whereas prior datasets typically lack the task components required to support these training stages.

%% file: Sec/4_method.tex
\section{Methodology}

\begin{table}[t]
\centering
\small
\caption{\textbf{Comparison with LLM-based analog circuit design methods.  
}We compare whether each method includes textual specifications, optimized sizing, incorporates simulation in the \textit{training} stage, trains the underlying model, supports multiple circuit types, and is open-source. Our method is the only one that supports all these dimensions. $\blacktriangle$ indicates that only the code is released as open source.}
\label{tab:method_comparison}
\resizebox{\linewidth}{!}{
\begin{tabular}{lcccccc}
\toprule
\textbf{Method} & \textbf{Text Specs} & \textbf{Opt. Sizing} & \textbf{Sim.-in-Training} & \textbf{Training} & \textbf{Multi-Type} & \textbf{Open-Source} \\
\midrule
Analogcoder~\cite{lai2025analogcoder}     & \checkmark & \xmark & -- & -- & \checkmark & $\blacktriangle$ \\
Analogcoder-Pro~\cite{lai2026analogcoder} & \checkmark & \checkmark & -- & -- & \checkmark & $\blacktriangle$ \\
AnalogSAGE~\cite{wang2025analogsage}      & \checkmark & \checkmark & -- & -- & \xmark & $\blacktriangle$ \\
AnalogGenie~\cite{gao2025analoggenie}     & \xmark & \xmark & \xmark & Pre-Training & \checkmark & \checkmark \\
LaMAGIC~\cite{chang2024lamagic}         & \xmark & \xmark & \xmark & SFT & \xmark & \xmark \\
Artisan~\cite{chen2024artisan}         & \checkmark & \checkmark & \xmark & SFT & \xmark & \xmark \\
AutoCircuit-RL~\cite{vijayaraghavan2025autocircuit}  & \checkmark & \xmark & \xmark & SFT + RL & \xmark & $\blacktriangle$ \\
\midrule
\textbf{Ours}   & \checkmark & \checkmark & \checkmark & SFT + RL & \checkmark & \checkmark \\
\bottomrule
\end{tabular}}
\end{table}
Tab.~\ref{tab:method_comparison} summarizes representative LLM-based methods for analog circuit design. Training-free methods improve the inference workflow around a frozen LLM, but not the LLM itself; the core design capabilities therefore remain test-time scaffolding rather than learned model capabilities. Moreover, reliance on external APIs can limit industrial autonomy and risk exposing proprietary circuit data, testbenches, and simulation pipelines. Training-based methods, on the other hand, adapt LLMs for analog circuit design, but many focus on unconditional topology synthesis rather than specification-conditioned analog circuit design. Others are restricted to specific circuit types or rely on simplified reward signals without directly using SPICE simulation outcomes. To fill this gap and progressively adapt LLMs to analog circuit design tasks, we propose a training pipeline consisting of three-stage SFT followed by simulation feedback-augmented RFT. 

\subsection{Three-Stage Supervised Fine-Tuning}

\begin{table}[t]
\centering
\caption{\textbf{Overview of the three-stage SFT.} \# denotes the number of training examples used.}
\label{tab:sft_stages}
\resizebox{\linewidth}{!}{
\small
\begin{tabular}{p{5.1cm}p{8.6cm}r}
\toprule
\textbf{SFT stage} & \textbf{Task Summary} & \textbf{\#} \\
\midrule
Circuit Knowledge Acquisition & Describe the circuit topology from a complete netlist. & 2,949 \\
Masked Component Prediction & Predict the missing component or connection in an incomplete netlist. & 1,967 \\
Simulation Feedback-Guided Correction & Revise a netlist using prior errors and simulation feedback. & 921  \\
\bottomrule
\end{tabular}}
\end{table}

Our preliminary experiments suggest that general-purpose open-source LLMs face two major challenges in analog circuit design: (1) they lack sufficient circuit-domain knowledge for netlist generation, and (2) they are not naturally equipped to revise failed designs using simulation feedback. We therefore design SFT as a three-stage curriculum that progressively builds the capabilities required for circuit topology design and feedback-based correction. The curriculum first develops circuit-domain knowledge, then strengthens component-level understanding through masked netlist completion, and finally teaches the model to revise failed designs using simulation feedback. Tab.~\ref{tab:sft_stages} summarizes the task and data size for each SFT stage. Example training instances are provided in App.~\ref{app:imple_details}.

\paragraph{Circuit Knowledge Acquisition.}
The first SFT stage aims to establish basic circuit-domain knowledge and familiarize the model with SPICE netlist syntax. We use the data annotated by Masala-CHAI~\cite{bhandari2024masala}, where the input is a SPICE-format netlist and the target output is the corresponding circuit name with a functional description. This enables the LLM to associate SPICE netlist representations with their corresponding circuit functionalities, providing a foundation for subsequent tasks.

\paragraph{Masked Component Prediction.}
The second SFT stage is designed to enhance the LLM's understanding of how circuit functionality depends on individual components and their connectivity. We intentionally mask one component at a time, covering its component type (e.g., NMOS, PMOS, resistor, or capacitor) and terminal connections, or entirely removing the full component. Given a partially masked netlist, the LLM is trained to predict the masked component. This task teaches the model to infer the functional role and proper connectivity of each component within a circuit.

\paragraph{Simulation Feedback-Guided Correction.}
Practical analog circuit design is inherently iterative: simulation results provide rich diagnostic information about whether a candidate circuit passes circuit topology check and satisfies target specifications, thereby guiding subsequent revisions. While the first two SFT stages provide the LLM with basic circuit-design capabilities, they do not explicitly train it to use simulation feedback for refining failed designs.
To develop this capability, we introduce a third SFT stage that teaches the model to correct failed designs based on simulation feedback. Each input consists of the original design task, an initial failed netlist generated by the model, and the corresponding simulation feedback from our simulation pipeline (see Fig.~\ref{fig:rl_pipeline} for details). The supervision is generated by a commercial closed-source teacher model (e.g., GPT-5.4) and includes failure analysis, feedback-informed modifications, and a corrected netlist for reference. By learning from these examples, the model learns to interpret simulation diagnostics and make feedback-driven revisions to failed designs.

\subsection{Simulation Feedback-Augmented Reinforcement Fine-Tuning}

While SFT provides an effective initialization, its imitation-based objective can constrain the LLM to reference design patterns and does not use simulation outcomes to directly optimize its generation policy. This is suboptimal for analog circuit design, where a candidate design must ultimately be validated by simulation. We therefore further optimize the LLM with RFT, using simulation outcomes as feedback signals beyond fixed reference demonstrations, which encourages generalization to unseen design examples. Specifically, we introduce simulation feedback-augmented RFT, which combines two training tasks: standard netlist generation and simulation feedback-guided correction. This enables the LLM to both generate initial designs and revise failed ones using simulation feedback.

\begin{figure}[t]
    \centering
    \resizebox{\textwidth}{!}{\includegraphics{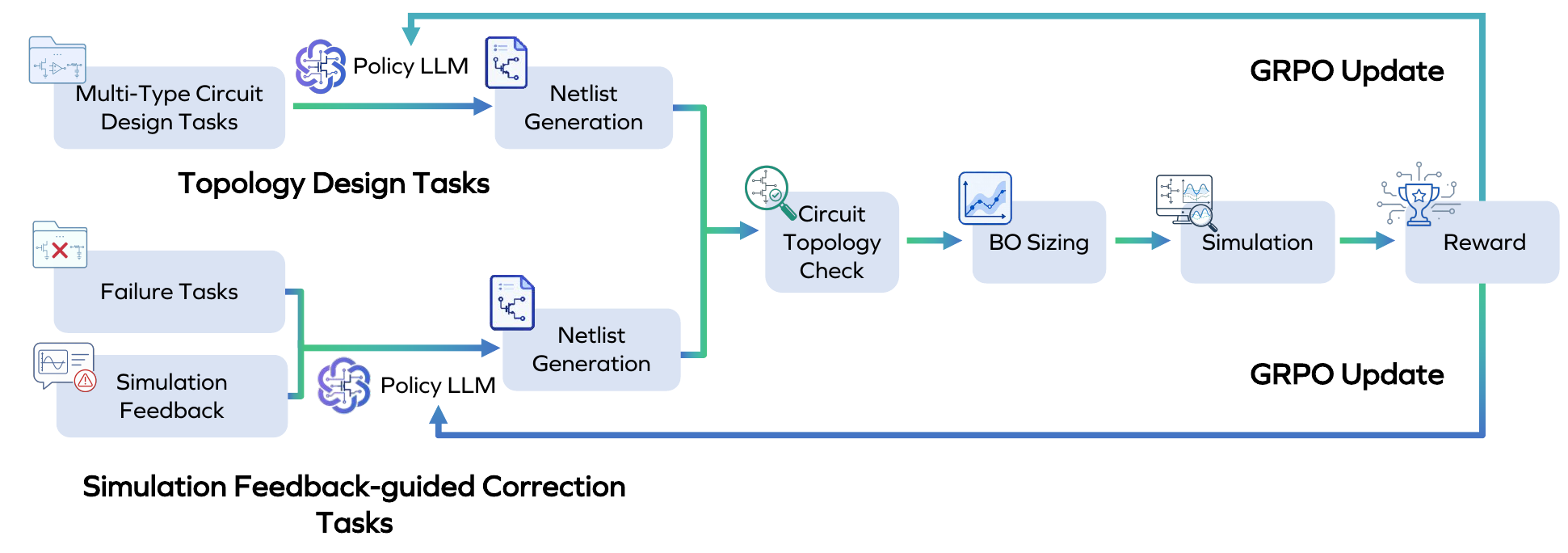}}
    \caption{\textbf{Overview of the simulation feedback-augmented RFT pipeline.} It combines standard circuit design tasks and simulation feedback-guided correction tasks. In both tasks, the generated netlist is evaluated through circuit topology check, BO sizing, and simulation. The resulting outcome is converted into a reward signal and used to update the policy with GRPO.}
    \label{fig:rl_pipeline}
\end{figure}

For both tasks, each netlist, either newly generated or revised from a failed design, undergoes the same simulation pipeline, including circuit topology check, BO sizing, and SPICE simulation. Circuits that fail the circuit topology check receive a reward of $-1$, circuits that satisfy all target specifications receive a reward of $1$, and all remaining cases receive a reward of $0$. We use GRPO~\cite{shao2024deepseekmath} to update the policy with these rewards, encouraging the model to generate structurally valid and specification-compliant circuit topologies. An illustration of the pipeline is provided in Fig.~\ref{fig:rl_pipeline}.

%% file: Sec/5_experiment.tex
\section{Experiments}

\subsection{Experimental Setup}
\label{sec:setup}

\paragraph{Implementation Details.}
We use Qwen3-8B~\cite{qwen3technicalreport} as the backbone model and train each of the three SFT stages for one epoch. For RFT, we use $128$ training examples per circuit type and train for $1600$ gradient steps. Detailed SFT and RFT hyperparameters are provided in App.~\ref{app:imple_details}.

We evaluate model performance under two practical analog circuit design settings. In the first setting, \textit{single-pass generation}, the model directly generates the netlist for a given design task from the textual description, labeled I/O ports, and target specifications. In the second setting, \textit{two-pass generation with simulation feedback}, success is measured after up to two inference rounds. The model first generates an initial netlist. If the initial design fails the circuit topology check, cannot be successfully simulated, or has unmet specifications, the model receives the failed netlist and simulation-guided feedback to produce a revised design. The test set contains $64$ instances for each circuit category, with no overlap with the training instances. For each test instance, we sample $8$ rollouts under each evaluation setting with temperature $0.2$, and compute pass@1 as the fraction of rollouts that successfully produce a valid design satisfying all target specifications. As baselines, we consider proprietary closed-source LLMs, including GPT-5.2~\cite{openai2025gpt52}, GPT-5.4~\cite{openai2026gpt54}, and Claude-Opus-4.7~\cite{anthropic2026opus47}, as well as larger open-source LLMs, including QwQ-32B~\cite{qwq32b}, Qwen3-32B~\cite{qwen3technicalreport}, and gpt-oss-20b~\cite{openai2025gptoss120bgptoss20bmodel}.

\subsection{Main Results}

\begin{table}[t]
\centering
\caption{\textbf{Pass rate comparison (\%) under \textit{single-pass generation.}} 
The model directly generates a complete circuit netlist from the design specification. 
We report pass@1 over $8$ rollouts per test instance. Bold indicates the best result in each column.}
\label{tab:single_pass_results}
\resizebox{\linewidth}{!}{
\begin{tabular}{llcccccc}
\toprule
\textbf{Model Type} & \textbf{Model} 
& \textbf{\makecell{Common\\Gate}} 
& \textbf{\makecell{Common\\Source}} 
& \textbf{\makecell{Current\\Mirror}} 
& \textbf{\makecell{Source\\Follower}} 
& \textbf{\makecell{Passive\\Mixer}} 
& \textbf{Average} \\
\midrule
\multirow{3}{*}{Closed-source}
& GPT-5.2 & 50.4 & 65.6 & 61.3 & 58.2 & 26.0 & 52.3 \\
& GPT-5.4 & 48.0 & 65.6 & 58.0 & 58.2 & 29.1 & 51.8\\
& Claude-Opus-4.7 & 45.7 & 65.6 & 62.7 & 63.7 & 39.8 & 55.5\\
\midrule
\multirow{4}{*}{Open-source}
& QwQ-32B & 41.0 & 63.5 & 27.9 & 35.2 & 25.6 & 38.6 \\
& Qwen3-32B & 42.0 & 55.3 & 38.3 & 34.9 & 29.3 & 40.0\\
& gpt-oss-20b & 42.2 & 57.2 & 7.8 & 34.6 & 16.8 & 31.7 \\
& \textbf{\method{}(Ours)} & \textbf{56.4} & \textbf{79.5} & \textbf{75.8} & \textbf{77.5} & \textbf{67.6} & \textbf{71.4} \\
\bottomrule
\end{tabular}}
\vspace{-4mm}
\end{table}

\begin{table}[t]
\centering
\caption{\textbf{Pass rate (\%) comparison under \textit{two-pass generation with simulation feedback.}}
The model first generates an initial circuit design and may revise it once using simulation-guided feedback.
We report the overall pass@1 after up to two generation attempts, where a rollout is counted as successful if either the initial design or the revised design satisfies all target specifications.
Bold indicates the best result in each column.}
\label{tab:two_pass_results}
\resizebox{\linewidth}{!}{
\begin{tabular}{llcccccc}
\toprule
\textbf{Model Type} & \textbf{Model} 
& \textbf{\makecell{Common\\Gate}} 
& \textbf{\makecell{Common\\Source}} 
& \textbf{\makecell{Current\\Mirror}} 
& \textbf{\makecell{Source\\Follower}} 
& \textbf{\makecell{Passive\\Mixer}} 
& \textbf{Average} \\
\midrule
\multirow{3}{*}{Closed-source}
& GPT-5.2 & 59.8 & 75.8 & 81.6 & 68.6 & 37.7 & 64.7 \\
& GPT-5.4 & 66.4 & 78.1 & 78.7 & 69.7 & 41.8 & 66.9 \\
& Claude-Opus-4.7 & 63.1 & 76.6 & 78.7 & 79.5 & 54.1 & 70.4\\
\midrule
\multirow{4}{*}{Open-source}
& QwQ-32B & 45.5 & 68.0 & 30.1 & 50.4 & 26.6 & 44.1 \\
& Qwen3-32B & 52.9 & 71.7 & 58.0 & 50.7 & 33.2 & 53.3\\
& gpt-oss-20b & 49.2 & 68.8 & 29.7 & 39.1 & 18.8 & 41.1 \\
& \textbf{\method{}(Ours)} & \textbf{67.8} & \textbf{81.8} & \textbf{82.8} & \textbf{80.1} & \textbf{78.5} & \textbf{78.2} \\
\bottomrule
\end{tabular}}
\vspace{-4mm}
\end{table}

\begin{figure}[h]
    \centering
    \includegraphics[width=0.8\textwidth]{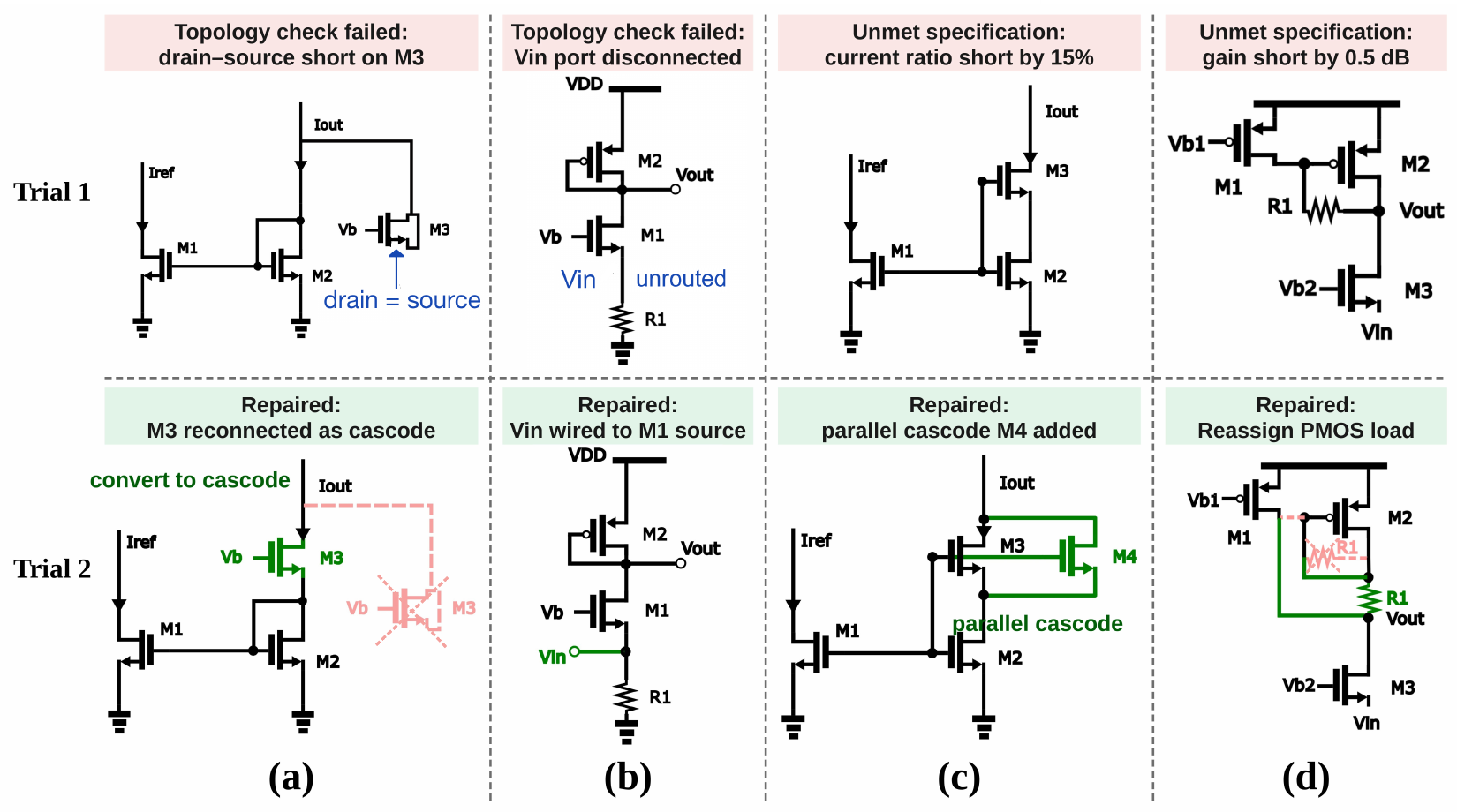}
    \caption{\textbf{Simulation-guided circuit revision examples.}
The top row shows initial failed designs, and the bottom row shows the corresponding revisions after incorporating simulation-augmented feedback. Cases (a) and (b) show structural recovery from topology-check failures, while cases (c) and (d) show metric-driven adjustment for specification violations. \textcolor{figred}{Red} marks removed parts, and \textcolor{figgreen}{green} marks newly introduced parts. Detailed information is provided in App.~\ref{app:case_study}.}
    \label{fig:qualitative_example}
        \vspace{-6mm}

\end{figure}

We report the main results in Tab.~\ref{tab:single_pass_results} and Tab.~\ref{tab:two_pass_results}. Overall, \method{} achieves the highest pass@1 across all five circuit types under both single-pass generation and two-pass generation with simulation-feedback. Notably, \method{} improves from single-pass to two-pass generation across all circuit types, showing that simulation feedback consistently helps the model revise failed designs. 
Under the single-pass setting,  
\method{} reaches an average pass@1 of 71.4\%, outperforming the strongest closed-source and open-source baselines by 15.9\% and 31.4\% respectively. This result shows that \method{} substantially improves the model's ability to generate valid analog circuit designs that satisfy all target specifications. With simulation-feedback correction, \method{} further improves its average pass@1 to 78.2\% and remains the best-performing model, exceeding the strongest closed-source and open-source baselines by 7.8\% and 24.9\% respectively. This improvement indicates that \method{} can effectively interpret simulation-guided diagnostics and revise failed designs. As case study, Fig.~\ref{fig:qualitative_example} shows $4$ representative cases. Detailed information is provided in App.~\ref{app:case_study}.

\paragraph{Out-of-Distribution Generalization}

To further evaluate whether the model generalizes beyond circuit types seen during training, we conduct an out-of-distribution (OOD) study on five unseen analog circuit types. As shown in Tab.~\ref{tab:OOD}, \method{} achieves the best or tied-best performance across all OOD circuit types under the two-pass generation setting, outperforming both the base model Qwen3-8B and the larger gpt-oss-20B model in most metrics. The improvement is particularly pronounced on active mixers, where \method{} raises pass@1 from 1.6\% to 41.4\% and pass@5 from 7.8\% to 85.0\% compared with the base model. \method{} also improves performance on structurally complex OOD circuit types, including fully differential amplifiers and delay cells. Overall, these results indicate that \method{} learns transferable analog circuit design priors beyond the circuit types included in training.
\vspace{-2mm}
\subsection{Ablation Study}

\paragraph{Ablation Study on Training Stage.}

\begin{figure}[th]
\vspace{-2mm}
    \centering
    \resizebox{\textwidth}{!}{\includegraphics{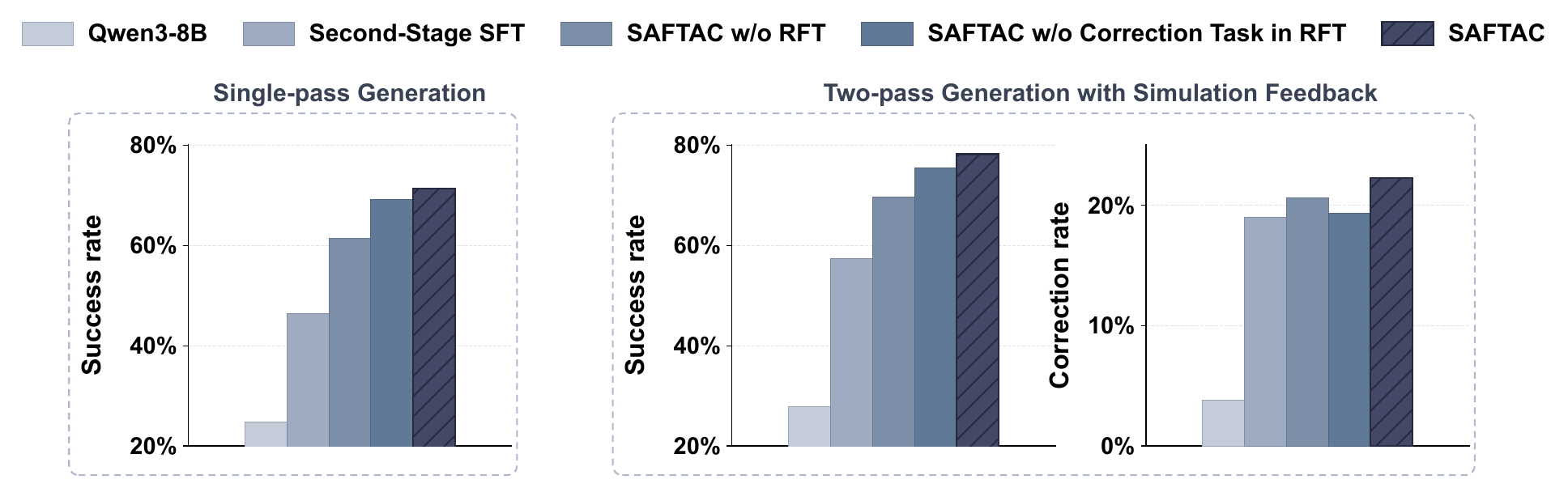}}
    \caption{\textbf{Ablation study on different training components in \method{}.} We compare different model variants under single-pass generation and two-pass generation with simulation feedback. The correction rate measures the fraction of initially failed designs that are successfully repaired in the second pass. Full \method{} achieves the best performance across all three metrics.}
    \label{fig:ablation_study}
\end{figure}

We include an ablation study to provide additional insight into the effect of different training stages in \method{}. As shown in Fig.~\ref{fig:ablation_study}, the base Qwen3-8B model shows limited performance, suggesting that general-purpose LLMs lack the domain-specific ability to reliably generate valid analog circuits. Adding the second-stage SFT improves single-pass generation, while the third-stage SFT provides additional gains in feedback-guided revision. The full \method{} achieves the strongest overall performance, reaching 71.4\% single-pass success rate, 78.2\% two-pass success rate, and 23.8\% feedback correction rate. Removing the correction task from RFT has a relatively small effect on single-pass performance but lowers the feedback correction rate. These trends suggest that staged SFT, RFT, and correction-task training provide complementary benefits.

\paragraph{Ablation Study on Reward Design.} To investigate whether partial reward design improves training effectiveness, we consider a partial reward that assigns a reward of $-1$ to circuits failing the topology check and otherwise assigns a reward equal to the fraction of satisfied target specifications. As shown in Tab.~\ref{tab:reward_ablation},
the original reward consistently outperforms the partial
reward across all five circuit types, improving the average
pass@1 by 9.6\% and 7.0\% under single-pass
and two-pass generation, respectively. These results suggest
that rewarding individual specification satisfaction does
not necessarily encourage designs that meet all target
specifications. A more principled partial reward may require identifying
partially valid circuit structures before assigning credit,
which remains challenging and is left for future work.

\begin{table}[h]
\centering
\caption{\textbf{Ablation study on reward design.}
We compare the original reward with an alternative
reward based on partial specification satisfaction.
We report pass@1 (\%) under single-pass generation and
two-pass generation with simulation feedback.
Bold indicates the best result in each column.}
\label{tab:reward_ablation}
\resizebox{\linewidth}{!}{
\begin{tabular}{llcccccc}
\toprule
\textbf{Setting} & \textbf{Reward Design}
& \textbf{\makecell{Common\\Gate}}
& \textbf{\makecell{Common\\Source}}
& \textbf{\makecell{Current\\Mirror}}
& \textbf{\makecell{Source\\Follower}}
& \textbf{\makecell{Passive\\Mixer}}
& \textbf{Average} \\
\midrule
\multirow{2}{*}{Single-pass}
& Partial Reward
& 46.5 & 70.7 & 61.3 & 68.4 & 62.1 & 61.8 \\
& Original Reward
& \textbf{56.4} & \textbf{79.5}
& \textbf{75.8} & \textbf{77.5}
& \textbf{67.6} & \textbf{71.4} \\
\midrule
\multirow{2}{*}{Two-pass}
& Partial Reward
& 57.8 & 76.6 & 72.5 & 73.4 & 75.6 & 71.2 \\
& Original Reward
& \textbf{67.8} & \textbf{81.8}
& \textbf{82.8} & \textbf{80.1}
& \textbf{78.5} & \textbf{78.2} \\
\bottomrule
\end{tabular}}
\end{table}

\paragraph{Generalizability Across Backbone Models.} To evaluate whether SAFTAC generalizes across different model architectures and scales, we apply SAFTAC to three additional open-source LLMs: Qwen3-4B\cite{qwen3technicalreport}, Llama-3.1-8B-Instruct\cite{grattafiori2024llama}, and GLM-4-9B\cite{glm2024chatglm}. As shown in Tab.~\ref{tab:backbone_generalization}, SAFTAC consistently improves performance across all backbones under both settings. Notably, applying SAFTAC to Llama-3.1-8B improves the average pass@1 from 23.6\% to 77.5\% under single-pass generation and from 28.7\% to 83.6\% with simulation feedback. These results suggest that SAFTAC's effectiveness is not restricted to the Qwen model family or a particular model scale.

\begin{table}[h]
\centering
\caption{\textbf{Generalizability of SAFTAC across backbone models.}
We apply the complete training pipeline to open-source LLMs
from different model families and scales.
We report the average pass@1 (\%) across five circuit types under single-pass generation and two-pass generation with simulation feedback.
Bold indicates the results after applying SAFTAC.}
\label{tab:backbone_generalization}
\resizebox{0.7\linewidth}{!}{
\begin{tabular}{lcccc}
\toprule
\multirow{2}{*}{\textbf{Backbone Model}}
& \multicolumn{2}{c}{\textbf{Single-pass Generation}}
& \multicolumn{2}{c}{\textbf{\makecell{Two-pass Generation\\with Simulation Feedback}}} \\
\cmidrule(lr){2-3}\cmidrule(lr){4-5}
& \textbf{Base} & \textbf{SAFTAC}
& \textbf{Base} & \textbf{SAFTAC} \\
\midrule
Qwen3-8B
& 27.4 & \textbf{71.4}
& 28.7 & \textbf{78.2} \\
Qwen3-4B
& 25.4 & \textbf{46.0}
& 30.6 & \textbf{57.7} \\
Llama-3.1-8B-Instruct
& 23.6 & \textbf{77.5}
& 28.7 & \textbf{83.6} \\
GLM-4-9B
& 26.5 & \textbf{54.5}
& 29.6 & \textbf{68.8} \\
\bottomrule
\end{tabular}}
\vspace{-2mm}
\end{table}

\begin{table}[]
\vspace{-2mm}
\caption{\textbf{OOD evaluation results on unseen circuit types.} 
We evaluate models on circuit types excluded from training and report pass@1 and pass@5 under \textit{two-pass generation with simulation feedback}. We sample 8 rollouts for each task. Bold indicates the best result in each column.}
\label{tab:OOD}
\resizebox{\columnwidth}{!}{%
\begin{tabular}{@{}ccccccccccc@{}}
\toprule
\multirow{2}{*}{\textbf{Model}} &
  \multicolumn{2}{c}{\textbf{Active Mixer}} &
  \multicolumn{2}{c}{\begin{tabular}[c]{@{}c@{}}\textbf{Differential Single-Ended}\\ \textbf{Amplifier}\end{tabular}} &
  \multicolumn{2}{c}{\begin{tabular}[c]{@{}c@{}}\textbf{Fully Differential} \\ \textbf{Amplifier}\end{tabular}} &
  \multicolumn{2}{c}{\begin{tabular}[c]{@{}c@{}}\textbf{Inverting} \\ \textbf{Delay Cell}\end{tabular}} &
  \multicolumn{2}{c}{\begin{tabular}[c]{@{}c@{}}\textbf{Non-Inverting} \\ \textbf{Delay Cell}\end{tabular}} \\ \cmidrule(l){2-11} 
         & \textbf{pass@1} & \textbf{pass@5} & \textbf{pass@1} & \textbf{pass@5} & \multicolumn{1}{c}{\textbf{pass@1}} & \multicolumn{1}{c}{\textbf{pass@5}} & \textbf{pass@1} & \textbf{pass@5} & \textbf{pass@1} & \textbf{pass@5} \\ \midrule
gpt-oss-20b & 4.7 & 20.1 & 2.7 & \textbf{12.6} & 44.1 & 80.8 & 8.6 & 35.1 & 0.0 & 0.0 \\
Qwen3-8B & 1.6     & 7.8     & 0.0     & 0.0     & 18.4                         & 37.7                         & 11.7     & 39.1     & 0.0     & 0.0     \\

\textbf{\method{}(Ours)}     & \textbf{41.4}    & \textbf{85.0}    & \textbf{3.1}     & \textbf{12.6}    & \textbf{58.6}                        & \textbf{88.1}                        & \textbf{16.4}     & \textbf{43.0}    & \textbf{6.3}    & \textbf{31.3}    \\ \bottomrule
\end{tabular}%
}
\end{table}

\subsection{Compatibility with Training-free Methods}

\begin{table}[th]
\centering
\caption{\textbf{Compatibility with the AnalogCoder inference pipeline.}
We evaluate different base models within the AnalogCoder~\cite{lai2025analogcoder} inference-time pipeline and report pass@1/pass@5 across circuit types. 
\method{} w/o AnalogCoder denotes direct generation without the AnalogCoder pipeline.}
\label{tab:compatibility_with_analogcoder}
\resizebox{\columnwidth}{!}{%
\begin{tabular}{@{}lccccccccc@{}}
\toprule
\multirow{2}{*}{\textbf{Base Model}} &
  \multicolumn{2}{c}{\textbf{Amplifier}} &
  \multicolumn{2}{c}{\textbf{Inverter}} &
  \multicolumn{2}{c}{\textbf{Current Mirror}} &
  \multicolumn{2}{c}{\textbf{Analog Adder}} \\
\cmidrule(lr){2-3}
\cmidrule(lr){4-5}
\cmidrule(lr){6-7}
\cmidrule(l){8-9}
& \textbf{pass@1} & \textbf{pass@5}
& \textbf{pass@1} & \textbf{pass@5}
& \textbf{pass@1} & \textbf{pass@5}
& \textbf{pass@1} & \textbf{pass@5} \\
\midrule
GPT-4o w/ AnalogCoder    & 100.0 & 100.0 & 100.0 & 100.0 & 56.7 & 78.6  & 100.0 & 100.0 \\
GPT-5.4 w/ AnalogCoder  & 100.0 & 100.0 & 100.0 & 100.0 & 56.7 & 78.6  & 93.3  & 100.0 \\
Qwen3-32B w/ AnalogCoder & 99.1  & 100.0 & 100.0 & 100.0 & 26.7 & 65.7  & 73.3  & 100.0 \\
QwQ-32B w/ AnalogCoder  & 60.0  & 99.0  & 100.0 & 100.0 & 60.0 & 95.8  & 93.3  & 100.0 \\
\midrule
\method{} w/o AnalogCoder & 98.1 & 100.0 & 100.0 & 100.0 & 76.7 & 100.0 & 66.7 & 100.0 \\
\method{} w/ AnalogCoder  & 100.0 & 100.0 & 100.0 & 100.0 & 76.7 & 100.0 & 73.3 & 100.0 \\
\bottomrule
\end{tabular}%
}
    \vspace{-4mm}
\end{table}

Training-free methods such as AnalogCoder~\cite{lai2025analogcoder} improve analog circuit generation through an inference-time feedback-enhanced workflow. Tab.~\ref{tab:compatibility_with_analogcoder} shows that \method{} can be effectively integrated into the AnalogCoder inference pipeline. First, combining \method{} with the AnalogCoder workflow further improves performance, indicating that our training-based approach is compatible with training-free inference-time correction methods. Second, when used as the base model in this pipeline, \method{} outperforms strong closed-source models and larger open-source models on circuit types included in our training data, such as amplifiers and current mirrors. In particular, it achieves 100\% pass@1 and pass@5 on amplifiers, and obtains the best pass@5 on current mirrors while matching the best pass@1 performance. Notably, \method{} also performs well on circuit types not included in training, including inverters and analog adders, reaching 100\% pass@5 for both categories.

\subsection{Failure Mode and Topology Analysis}
\label{sec:failure_analysis}

\textbf{Failure Mode Analysis.}
To better understand where SAFTAC's performance improvements
originate, we categorize generated circuits into five outcomes: syntactically invalid netlists, topology
check failures, BO sizing failures, specification violations,
and successful designs. Detailed classification criteria are provided in App.\ref{app:failure}. As shown in Tab.~\ref{tab:failure_modes}, general-purpose open-source LLMs exhibit substantial syntax failure rates, whereas SAFTAC produces no syntactically invalid netlists. In contrast, strong closed-source models rarely generate invalid netlists but frequently fail to satisfy the target specifications. These results suggest that SAFTAC improves both structural validity and specification satisfaction.

\textbf{Topology Diversity and Correction Behavior.}
We further analyze the structural diversity of successful
designs. As shown in Tab.~\ref{tab:topology_analysis},
SAFTAC generates the largest number of unique successful circuit topologies among the evaluated models.
Furthermore, SAFTAC largely preserves circuit complexity
during correction. In comparison, Claude-Opus-4.7 and
GPT-5.4 increase their average device counts from 3.59
to 4.09 and from 4.70 to 5.06, respectively.
These observations suggest that SAFTAC tends to make
targeted modifications while preserving existing circuit
structures, whereas general-purpose models more frequently
expand circuit complexity during revision.
\vspace{-2mm}

%% file: Sec/6_conclusion.tex
\section{Conclusion}
\vspace{-2mm}
We introduced \method{}, a simulation-augmented fine-tuning framework for analog circuit design. To support training, we construct a large-scale simulation-grounded dataset with complete task information across diverse circuit types. We hope this work provides a foundation for training open-source LLMs as practical assistants for simulation-grounded analog circuit design. 

\section*{Acknowledgment}
This material is based upon work supported by the National Science Foundation under Grant No. CNS 2137288 - Center for Advanced Electronics through Machine Learning (CAEML) and its industry members. We would also like to thank Chris Cheng, Chinq-Shiun (Gary) Chiu, and Priyank Kashyap for their valuable discussions and feedback.

%% file: Sec/7_appendix.tex
\newpage
\appendix

\section{Discussion}

\subsection{Limitations and Future Work}
\label{sec:app_limitation}
Our feedback-guided evaluation currently follows a two-pass setting, where the model is allowed to revise each failed design only once. More iterative feedback loops may further improve difficult cases. Additionally, our current framework relies on a single model to perform both initial netlist generation and feedback-guided revision. A promising direction is to use our simulation-grounded data pipeline to construct specialized datasets, including circuit topology check data, BO sizing trajectories, simulation-result data, and correction traces, and train multiple expert models that collaborate in a multi-agent analog circuit design system. Additionally, given the high cost of BO sizing and SPICE simulation, another promising direction is to improve the data efficiency of simulation-grounded RFT by leveraging recent advances in data-efficient RFT~\cite{sun2026improving,tang2026towards,nguyen2026adaptive}.

\subsection{Additional Related Work}

\paragraph{AI-assisted device sizing.}
Device sizing is a fundamental step in analog circuit design, where a fixed topology is tuned by selecting continuously-valued device parameters to satisfy target performance specifications under simulation. While traditional methods rely on analytical equations, expert-crafted design heuristics, or black-box optimization, recent studies have introduced AI and ML techniques to improve sizing efficiency and transferability. For example, GCN-RL~\cite{wang2020gcn} represents circuits as graphs and combines graph neural networks with reinforcement learning for transferable transistor sizing. Actor-critic optimization frameworks~\cite{dutta2026can} use learned proposal and evaluation modules to guide analog design optimization. INSIGHT~\cite{poddar2025insight} employs transformers as neural simulators for efficient performance prediction. These methods can improve sample efficiency and reduce the human effort required for sizing. However, since they typically assume a predefined circuit type or even a fixed netlist, they are not directly applicable to open-ended analog circuit design task.

\paragraph{Search-Based Topology Synthesis.}
Before the recent adoption of LLM-based methods, automated analog circuit design was commonly formulated as a discrete search problem over candidate circuit structures. Representative methods include genetic-programming frameworks~\cite{mcconaghy2011trustworthy}, tree-structured construction methods for generating candidate topologies~\cite{zhao2020automated}, functional-block composition for operational-amplifier synthesis~\cite{abel2022fuboco}, and reinforcement-learning methods for constraint-driven topology exploration~\cite{zhao2024automated,zhao2022analog}.
These methods reduce the search space by embedding expert knowledge through handcrafted rules, predefined subcircuits, functional blocks, or fast feasibility filters. However, each new performance specification typically requires iterative candidate generation, evaluation, and sizing, often with many simulation or surrogate-evaluation queries. As a result, although search-based methods can be effective within constrained circuit families, their scalability and adaptability remain limited when moving toward diverse, specification-conditioned analog circuit design tasks.


\section{Additional Preliminaries}

\subsection{Circuit Type}
\label{app:circuit_type}

As summarized in Tab.~\ref{tab:dataset_statistics}, our datasets covers a representative spectrum of analog circuit types, ranging from basic building blocks such as common gate amplifiers, common source amplifiers, source followers, and current mirrors to more structured circuits such as differential single-ended amplifiers, fully differential amplifiers, delay cells, passive mixers and active mixers. Together, these circuit types cover complementary analog design challenges, including local signal-path construction and biasing in single-transistor stages, reference-current generation and output-voltage compliance in current mirrors, stage composition and symmetry preservation in amplifiers, transient oscillation behavior in delay cells, and frequency translation with port isolation in mixers. In total, the datasets contains \textbf{8,626} samples across \textbf{10} circuit types.

\begin{table}[t]
\centering
\caption{\textbf{Dataset statistics by circuit type.}
We report the number of unique reference netlists and
specification-conditioned design tasks for each circuit type.
Multiple design tasks can be derived from the same reference
netlist with different target specifications.
Performance specifications are color-coded by optimization
direction: red indicates higher-is-better, green indicates
lower-is-better, and black indicates equality constraints.
See App.~\ref{app:specs} for specification definitions.}
\label{tab:dataset_statistics}
\resizebox{\linewidth}{!}{
\begin{tabular}{lccc}
\toprule
\textbf{Circuit Type}
& \textbf{Performance Specs}
& \textbf{\makecell{\# Unique\\Reference Netlists}}
& \textbf{\makecell{\# Design\\Tasks}} \\
\midrule

Common Gate Amplifier &
\textcolor{red}{gain}, \textcolor{red}{bandwidth}, \textcolor{green}{power} &
15 & 1,372 \\

Common Source Amplifier &
\textcolor{red}{gain}, \textcolor{red}{bandwidth}, \textcolor{green}{power} &
18 & 1,861 \\

Source Follower &
\textcolor{red}{gain}, \textcolor{red}{bandwidth}, \textcolor{green}{power} &
8 & 720 \\

Current Mirror &
\textcolor{red}{$r_{out}$}, \textcolor{red}{$V_{out,\max}$}, \textcolor{green}{$V_{out,\min}$}, \textcolor{black}{current ratio} &
12 & 478 \\

Differential Single-Ended Amplifier &
\textcolor{red}{gain}, \textcolor{red}{bandwidth}, \textcolor{red}{output swing}, \textcolor{red}{phase margin}, \textcolor{red}{CMRR}, \textcolor{green}{power} &
23 & 955 \\

Fully Differential Amplifier &
\textcolor{red}{gain}, \textcolor{red}{bandwidth}, \textcolor{red}{output swing}, \textcolor{red}{phase margin}, \textcolor{red}{CMRR}, \textcolor{green}{power} &
11 & 1,632 \\

Inverting Delay Cell &
\textcolor{red}{$V_{pp}$}, \textcolor{green}{power}, \textcolor{black}{frequency}, \textcolor{black}{duty cycle} &
11 & 519 \\

Non-Inverting Delay Cell &
\textcolor{red}{$V_{pp}$}, \textcolor{green}{power}, \textcolor{black}{frequency}, \textcolor{black}{duty cycle} &
2 & 109 \\

Passive Mixer &
\textcolor{red}{$\mathrm{IIP3}$}, \textcolor{red}{conversion gain}, \textcolor{green}{LO leakage}, \textcolor{green}{RF leakage} &
6 & 635 \\

Active Mixer &
\textcolor{red}{$\mathrm{IIP3}$}, \textcolor{red}{conversion gain}, \textcolor{green}{LO leakage}, \textcolor{green}{RF leakage}, \textcolor{green}{power} &
8 & 345 \\

\midrule
\textbf{Total} & \textemdash & \textbf{114} & \textbf{8,626} \\
\bottomrule
\end{tabular}}
\end{table}

\paragraph{Common Gate Amplifier.}
A common gate amplifier is a single-transistor gain stage in which the input is applied at the source, the gate is held at a bias voltage, and the output is taken from the drain. This type tests whether the model can construct the correct source-input and gate-bias connectivity for a low-input-impedance, non-inverting small-signal stage. It is evaluated using gain, bandwidth, and power.

\paragraph{Common Source Amplifier.}
A common source amplifier is a single-transistor gain stage in which the input drives the gate and the output is taken from the drain. It represents the canonical inverting voltage-gain configuration used as a local gain stage in many larger analog circuits. This type tests whether the model can create a valid gain path, bias the transistor in a useful operating region, and trade off gain, bandwidth, and power.

\paragraph{Source Follower.}
A source follower, or common drain stage, is a single-transistor buffer in
which the input drives the gate and the output is taken from the source. It
provides near-unity voltage gain, high input impedance, and low output
impedance. This type tests whether the model can preserve the correct
source-output connectivity and produce a valid DC level shift while satisfying gain, bandwidth, and power constraints.

\paragraph{Current Mirror.}
A current mirror is a biasing cell that copies a reference current into one or more output branches with a specified current ratio. In our dataset, current mirrors test whether the model can form valid diode-connected reference devices, output mirror branches, and optional cascode devices. The associated specifications evaluate output resistance, output-voltage compliance, and mirrored current ratio.

\paragraph{Differential Single-Ended Amplifier.}
This type denotes an amplifier that compares two input signals and produces one output signal. The circuit usually relies on a matched input branch to sense the input difference, a biasing branch to provide the operating current, and a load branch to convert the sensed difference into an output voltage. This circuit type evaluates whether the model can construct the correct differential-to-single-ended signal path, provide a valid biasing structure, and meet the target gain, bandwidth, output swing, phase margin, common-mode rejection ratio, and power specifications.

\paragraph{Fully Differential Amplifier.}
This type denotes an amplifier that receives a pair of differential input signals and produces a pair of differential output signals. A typical design contains two balanced output branches, so the circuit must preserve symmetry between the positive and negative signal paths while maintaining proper biasing. In addition, the output voltages must be kept around a desired common-mode level, either through a dedicated common-mode control structure or an equivalent biasing mechanism. This circuit type evaluates whether the model can construct a balanced differential signal path, maintain output symmetry, control the output common-mode level, and satisfy the required gain, bandwidth, output swing, phase margin, common-mode rejection ratio, and power specifications.



\paragraph{Inverting Delay Cell.}
This type denotes an inverting delay cell evaluated using a ring-oscillator testbench. The generated circuit is a single delay stage, and during evaluation, three identical instances are connected to form a closed loop. Because the loop contains an odd number of inversions, it can sustain oscillation without an external periodic input. The resulting three-stage ring oscillator is evaluated by peak-to-peak output voltage, oscillation frequency, duty cycle, and power.

\paragraph{Non-Inverting Delay Cell.}
This type denotes a non-inverting delay cell, typically with differential input and differential output, evaluated using a ring-oscillator testbench. During evaluation, three identical instances of the generated delay stage are connected into a closed loop, and the final feedback path is cross-coupled to provide the phase inversion required for oscillation. The resulting three-stage ring oscillator is evaluated by peak-to-peak output voltage, oscillation frequency, duty cycle, and power.

\paragraph{Passive Mixer.}
A passive mixer converts a radio-frequency input signal to an intermediate-frequency output by using transistor switches driven by a local-oscillator signal. The local oscillator periodically turns the switches on and off, which shifts the input signal to the desired output frequency band. Since the transistors mainly act as switches rather than biased amplifying devices, the mixer core ideally consumes little static power. This circuit type evaluates correct port connectivity and frequency conversion, with specifications on conversion gain, third-order input intercept point, local-oscillator leakage, and radio-frequency leakage.

\paragraph{Active Mixer.}
An active mixer converts a radio-frequency input signal to an intermediate-frequency output while also providing signal amplification. A typical design first uses a biased input stage to convert the input voltage signal into a current signal, and then uses local-oscillator-controlled switching devices to steer this current to the output at the desired translated frequency. Compared with a passive mixer, an active mixer can provide positive conversion gain, but it requires proper biasing and consumes static power. This circuit type evaluates whether the model can combine input transconductance generation, local-oscillator switching, biasing, and radio-frequency/intermediate-frequency port isolation, while satisfying conversion gain, third-order input intercept point, local-oscillator leakage, radio-frequency leakage, and power specifications.

\subsection{Performance Specifications}
\label{app:specs}

We use the following performance specifications in
Table~\ref{tab:dataset_statistics}. Each specification is associated with an
optimization direction: \textcolor{red}{red} indicates a higher-is-better
constraint, \textcolor{green}{green} indicates a lower-is-better
constraint, and black indicates an equality target.

\paragraph{Gain.}
Gain is the small-signal voltage gain extracted from AC analysis. For
single-ended circuits, it is reported as the magnitude of the output-to-input
voltage transfer function. For fully-differential circuits, gain refers to the
differential-mode gain. Higher gain indicates stronger voltage amplification
and is therefore treated as a higher-is-better specification.

\paragraph{Bandwidth.}
Bandwidth is the small-signal $-3\,\mathrm{dB}$ bandwidth extracted from the AC
response. It is defined as the frequency at which the gain magnitude drops by
$3\,\mathrm{dB}$ from its low-frequency value. A larger bandwidth indicates
that the circuit can preserve its intended gain over a wider frequency range.

\paragraph{Power.}
Power is the supply power consumed by the circuit under the corresponding
testbench condition. For DC or small-signal amplifier tasks, it is computed
from the supply current at the operating point. For transient or periodically
driven tasks, such as ring oscillators and mixers, it is measured as the
average supply power over the relevant steady-state simulation window. Lower
power is preferred.

\paragraph{Output Swing.}
Output swing is the usable peak-to-peak output-voltage range measured from a
transient or closed-loop testbench. For fully-differential circuits, the swing
is evaluated on the differential output. A larger output swing generally
indicates a larger allowable signal range before clipping or severe waveform
distortion.

\paragraph{Phase Margin.}
Phase margin characterizes the stability of an amplifier under feedback. It is
measured from the open-loop or loop-gain response at the unity-gain crossing:
\[
\mathrm{PM} = 180^{\circ} + \angle A(f_u),
\]
where $f_u$ is the unity-gain frequency. A larger phase margin, reported in
degrees, indicates a more stable and less under-damped feedback response.

\paragraph{CMRR.}
Common-mode rejection ratio (CMRR) measures the ability of a differential
amplifier to reject input signals that appear equally on both input terminals.
It is defined as
\[
\mathrm{CMRR} = 20\log_{10}\left(\frac{A_{dm}}{A_{cm}}\right),
\]
where $A_{dm}$ is the differential-mode gain and $A_{cm}$ is the common-mode
gain. Higher CMRR, reported in dB, indicates stronger rejection of common-mode
disturbances.

\paragraph{$r_{out}$.}
Output resistance is used for current-mirror tasks. It is the small-signal
resistance looking into the mirror output node and is extracted by perturbing
the output voltage and measuring the corresponding change in output current:
\[
r_{out} = \frac{\partial V_{out}}{\partial I_{out}}.
\]
A higher output resistance indicates a more ideal current-source behavior.

\paragraph{$V_{out,\max}$ / $V_{out,\min}$.}
The current-mirror output-voltage range is described by an upper and lower
compliance limit. $V_{out,\max}$ is the largest output voltage over which the
mirror remains within the required current accuracy, while $V_{out,\min}$ is
the minimum output voltage needed to maintain the programmed current within
the specified tolerance. Therefore, a larger $V_{out,\max}$ and a smaller
$V_{out,\min}$ are preferred.

\paragraph{Current Ratio.}
Current ratio is the ratio between the mirrored output current and the
reference current:
\[
\mathrm{current\ ratio} = \left|\frac{I_{out}}{I_{ref}}\right|.
\]
It is specified as an equality target because the goal is not simply to
maximize or minimize the mirrored current, but to match the requested mirror
ratio.

\paragraph{$V_{pp}$.}
$V_{pp}$ is the peak-to-peak voltage of the steady-state oscillation waveform
in the ring-oscillator testbench. It is extracted after the startup transient
has decayed. A larger $V_{pp}$ indicates a stronger and more rail-to-rail
oscillation.

\paragraph{Frequency.}
Frequency is the oscillation frequency of the ring oscillator constructed from
the generated delay cell. It is measured from the steady-state output waveform,
typically using zero crossings or period extraction after startup. It is
treated as an equality target because each task specifies a desired
oscillation frequency.

\paragraph{Duty Cycle.}
Duty cycle is the fraction of one oscillation period during which the output
waveform stays above the chosen switching threshold, typically the mid-supply
level. A symmetric oscillator targets a duty cycle near $50\%$, so this metric
is treated as an equality target.

\paragraph{IIP3.}
Third-order input-referred intercept point (IIP3) measures mixer linearity. It
is extracted from a two-tone RF test in which the fundamental IF output tones
and the third-order intermodulation products are measured. Higher IIP3
indicates better large-signal linearity and lower third-order distortion.

\paragraph{Conversion Gain.}
Conversion gain measures the amplitude transfer from the RF input to the IF
output under nominal LO drive. It is typically expressed in dB. Passive mixers
typically exhibit negative values, i.e., conversion loss, while active mixers
can provide positive conversion gain. Higher conversion gain is preferred.

\paragraph{LO Leakage / RF Leakage.}
LO leakage and RF leakage measure unwanted spectral components at the IF
output that originate from feedthrough of the LO or RF signal. These metrics
quantify port-to-port isolation in the mixer. Lower leakage indicates better
isolation and is therefore preferred.

\subsection{Testbench Construction and Validation}
\label{app:testbench}

To ensure reliable and reproducible circuit evaluation,
we manually design and validate testbench templates for
each circuit type. Given a design task, deterministic
Python scripts instantiate the corresponding template
based on its circuit type, I/O ports, target specifications,
and external load conditions. Therefore, all tasks of the
same circuit type share consistent simulation procedures
and performance extraction logic. We describe the
testbench implementations for different circuit types
below.

\paragraph{Single-Input, Single-Output Amplifiers.}
Common source amplifiers, common gate amplifiers,
and source followers share a single-ended amplifier
testbench. A tunable DC bias and a unit-amplitude AC
small-signal excitation are applied at the input,
while the specified capacitive or resistive load
is connected to the output. DC operating-point
analysis and AC frequency sweeps are performed
to extract low-frequency gain, $-3\,\mathrm{dB}$
bandwidth, and power consumption.

\paragraph{Differential-Input, Single-Ended Amplifiers.}
For differential-input amplifiers, direct open-loop
simulation may drive the output into saturation or
prevent a stable DC operating point due to their
high open-loop gain. Physically breaking the feedback
loop can also alter the original biasing and loading
conditions. We therefore employ unity-gain feedback
to establish a stable operating point and use Tian's
method to extract the loop gain without physically
opening the loop. Specifically, since NGspice does
not provide an equivalent to the \texttt{stb}
analysis in Cadence Spectre, we perform series
voltage injection and parallel current injection
at the same loop-probe point. The resulting AC
responses are combined to reconstruct the return
ratio $T(f)$. Under unity feedback ($\beta=1$),
the reconstructed loop gain corresponds to the
amplifier's open-loop gain. We then extract
low-frequency gain, $-3\,\mathrm{dB}$ bandwidth,
unity-loop-gain frequency, and phase margin.
Output swing, CMRR, and power are obtained through
DC sweeps, differential- and common-mode AC analyses,
and supply-current measurements at the closed-loop
operating point.

\paragraph{Fully Differential Amplifiers.}
For fully differential amplifiers, an ideal
common-mode feedback loop stabilizes the output
common-mode voltage, and identical external loads
are connected to both outputs. We measure the
differential loop gain through voltage injection
at an ideal balun-based cut point implemented
using controlled sources. Since the driving-point
impedance toward the source is zero at this point,
voltage injection alone is sufficient without
additional current injection. The resulting
responses are used to extract gain, bandwidth,
phase margin, differential output swing, CMRR,
and power consumption.

\paragraph{Current Mirrors.}
All current mirrors share a DC output-voltage
sweep testbench. The output current is measured
through a series test voltage source, from which
we calculate the current ratio and valid output-voltage
range. The valid range is defined as the continuous
region where the output current remains within
$\pm 5\%$ of its nominal value. Output resistance
is measured only when this region is sufficiently
wide and the output current is nonzero, preventing
invalid measurements at improper operating points.

\paragraph{Oscillators.}
Inverting and non-inverting delay cells share
a transient simulation and waveform analysis
procedure, with different ring interconnections.
For delay cells with explicit input and output
ports, the testbench automatically constructs
an $N$-stage ring oscillator according to the
specified configuration, handling single-ended,
differential, and feedback-polarity settings.
Fully connected ring or relaxation oscillators
are simulated directly. A small initial
perturbation is applied to initiate oscillation,
and the steady-state waveform is analyzed to
extract oscillation status, frequency, duty
cycle, and power consumption. Failed simulations,
insufficient oscillation amplitude, or too few
detected cycles are classified as non-oscillating
rather than assigned invalid frequency values.

\paragraph{Mixers.}
Active and passive mixers share a port-driving
and spectral-analysis framework. Based on the
RF, LO, and IF port definitions, the testbench
configures single-ended or differential excitation,
RF source impedance, large-signal LO input, and
IF loading conditions. Single-tone transient
simulations with windowed FFT analysis are used
to measure conversion gain, RF-to-IF leakage,
LO-to-IF leakage, and average power consumption.
Two-tone simulations are performed to extract
IIP3 from the fundamental and third-order
intermodulation components. Frequency matching,
local noise-floor checks, and multi-amplitude
consistency checks are applied to reject
unreliable measurements caused by spectral
leakage, noise, or compression.

\paragraph{Testbench Validation.}
All testbench templates and specification
extraction procedures are manually reviewed
by circuit experts. In addition, reference
circuits from instructor-developed course
materials provide established testbenches
previously validated in Cadence. We translate
these testbenches to NGspice and compare the
extracted performance metrics across the two
simulators to verify the correctness of the
conversion, simulation setup, and measurement
procedures. These reference circuits also serve
as sanity-check cases for topology checking,
BO-based sizing, and circuit simulation.

\begin{tcolorbox}[title={Stage 1 (Circuit Knowledge Acquisition) SFT Example Training Instance}]
\textbf{\texttt{\{"question":}}

\quad \texttt{"What is the function of the following circuit described by this netlist?}

\medskip
\quad \texttt{Netlist:}

\begin{lstlisting}[basicstyle=\ttfamily\small, breaklines=true]
    subckt DUT VDD VSS VB1 VIN1 VOUT1
    R1 (net13 VB1) resistor
    R0 (VDD VOUT1) resistor
    M0 (VOUT1 net13 VSS VSS) nmos4
    C0 (net13 VIN1) capacitor
    ends DUT
\end{lstlisting}

\textbf{\texttt{"output":}}

\quad The schematic shows a common-source MOSFET amplifier. Transistor \(M_0\) is biased through resistor \(R_1\) to \(V_{\mathrm{B1}}\), and the input signal \(V_{\mathrm{IN1}}\) is AC-coupled to the gate through capacitor \(C_0\). Resistor \(R_0\) acts as the load between \(V_{\mathrm{DD}}\) and \(V_{\mathrm{OUT1}}\), where the amplified output is taken.

\textbf{\texttt{\}}}
\end{tcolorbox}

\begin{center}

\begin{tcolorbox}[breakable, title={Stage 2 (Masked Component Prediction) SFT Example Training Instance}]
\textbf{\texttt{\{"question":}}

\quad \texttt{"Complete the incomplete SPICE netlist with the easiest answer.}

\medskip
\quad \texttt{Here is an example:}

    \begin{lstlisting}[basicstyle=\ttfamily\small, breaklines=true]
    Incomplete Netlist:
    
    subckt DUT VDD VSS vin vout
    N0 (VDD vin vout VSS) tsmc18dN w=Wu l=Ln
    [MASK] (vout VSS) [MASK] r=RK m=1
    V0 (vin VSS) vsource type=dc dc=Vm
    ends DUT
    
    Task Analysis:
    
    We are told it is a common-drain (source follower) and the device pin order is D-G-S-B. The NMOS is already wired as D->VDD, G->vin , S->vout, B->VSS, which is correct for a CD stage. A CD needs a load from vout to VSS for DC bias; therefore the missing part is the resistor. We should add 'R0' to the first MASK and add 'resistor' to the second MASK.
    
    So the complete netlist is:
    
    subckt DUT VDD VSS vin vout
    N0 (VDD vin vout VSS) tsmc18dN w=Wu l=Ln
    R0 (vout VSS) resistor r=RK m=1
    V0 (vin VSS) vsource type=dc dc=Vm
    ends DUT
    \end{lstlisting}

\quad \texttt{Here is another example:}

    \begin{lstlisting}[basicstyle=\ttfamily\small, breaklines=true]
    Incomplete Netlist:
    
    subckt DUT VDD VSS vin vout
    [MASK] (VDD vin vout VSS) [MASK] w=Wu l=Ln
    R0 (vout VSS) resistor r=RK m=1
    V0 (vin VSS) vsource type=dc dc=Vm
    ends DUT
    
    Task Analysis:
    We are told it is a common-drain stage. The resistor already provides the required return path vout -> VSS. The missing active device is the NMOS wired as D->VDD, G->vin, S->vout, B->VSS. We should add 'N0' to the first MASK and add 'tsmc18dN' to the second MASK.
    
    So the complete netlist is:
    
    subckt DUT VDD VSS vin vout
    N0 (VDD vin vout VSS) tsmc18dN w=Wu l=Ln
    R0 (vout VSS) resistor r=RK m=1
    V0 (vin VSS) vsource type=dc dc=Vm
    ends DUT
    \end{lstlisting}

\quad \texttt{Here are some rules you should follow all the time:}

    \begin{lstlisting}[basicstyle=\ttfamily\small, breaklines=true]
    1. Do NOT add dot before subckt or ends
    2. Use these EXACT parameters for ALL components:
       - Resistors: r=RK (not actual values)
       - Capacitors: c=CK (not actual values)
       - NMOS transistors: w=Wn l=Ln
       - PMOS transistors: w=Wu l=Ln
       - Technology: tsmc18dN (NMOS), tsmc18dP (PMOS)
       - Process: 180nm TSMC
    3. The first [MASK] should be filled with the component name
    4. The second [MASK] should be filled with the component type (NMOS should be tsmc18dN, PMOS should be tsmc18dP, other choice includes resistor and capacitor)
    5. For the mosfet components, name it as N/P+integer based on its type(tsmc18dN tsmc18dP)
    6. Only modify the [MASK], do NOT modify the existing part of the netlist
    8. Maintain all existing net names when connecting the missing component
    \end{lstlisting}

\quad \texttt{Here is the problem:}

    \begin{lstlisting}[basicstyle=\ttfamily\small, breaklines=true]
    Circuit Type:
    single_balanced_passive_mixer
    
    Requirements:
    Differential passive mixer with two NMOS switches (M0/M1). Single-ended RF input, differential LO (VLO1/VLO2), differential output (VOUT1/VOUT2). Resistive loads to VSS, AC coupling capacitors.
    
    Incomplete Netlist:
    subckt DUT VSS VOUT1 VOUT2 VRF1 VLO1 VLO2
    C1 (VLO1 VRF1) capacitor c=CK
    [MASK] (VOUT1 VLO1) [MASK] c=CK
    C2 (VOUT2 VLO2) capacitor c=CK
    C3 (VLO2 VRF1) capacitor c=CK
    N1 (VOUT2 VLO2 VRF1 VSS) tsmc18dN w=Wu l=Ln
    N0 (VOUT1 VLO1 VRF1 VSS) tsmc18dN w=Wu l=Ln
    ends DUT
    \end{lstlisting}

\quad \texttt{First give a short task analysis, then return the complete netlist, starting with subckt and ending with ends."}

\medskip
\textbf{\texttt{"output":}}

\quad Task analysis: A single-balanced passive mixer uses two NMOS switches (already present as N0/N1) driven by differential LO (VLO1/VLO2) and a single-ended RF input (VRF1), producing differential outputs (VOUT1/VOUT2). The missing element should satisfy the AC coupling capacitors requirement; the incomplete line is a capacitor from VOUT1 to its LO-related node VLO1, symmetric with the other capacitive connections shown. So fill the missing component as a capacitor instance name and type.

    \begin{lstlisting}[basicstyle=\ttfamily\small, breaklines=true]
    subckt DUT VSS VOUT1 VOUT2 VRF1 VLO1 VLO2
    C1 (VLO1 VRF1) capacitor c=CK
    C0 (VOUT1 VLO1) capacitor c=CK
    C2 (VOUT2 VLO2) capacitor c=CK
    C3 (VLO2 VRF1) capacitor c=CK
    N1 (VOUT2 VLO2 VRF1 VSS) tsmc18dN w=Wu l=Ln
    N0 (VOUT1 VLO1 VRF1 VSS) tsmc18dN w=Wu l=Ln
    ends DUT
    \end{lstlisting}
    
\textbf{\texttt{\}}}
\end{tcolorbox}

\end{center}

\begin{center}
\begin{tcolorbox}[breakable, title={Stage 3 (Simulation Feedback-Guided Correction) SFT Example Training Instance}]
\textbf{\texttt{\{"question":}}

\begin{lstlisting}[basicstyle=\ttfamily\small, breaklines=true, columns=fullflexible, keepspaces=true,escapeinside={(*@}{@*)}]
# Feedback Correction Task

## Objective
You are correcting a previously generated circuit netlist.
Your goal is **not** to redesign from scratch by default. First determine the failure type, then apply the **smallest topology edit** that can realistically fix the failure.

You will be given:
- the original task specifications
- the previous incorrect netlist
- a structured failure category
- a concise metric summary
- a diagnosis summary
- optional operating-point observations
- optional BO/debug summary

## Priority Order
Follow this decision order strictly:
1. **Legality**: fix floating nodes, unused ports, impossible DC paths, wrong terminal roles, broken source/drain connections.
2. **Topology consistency**: make sure the circuit matches the named topology and topology description.
3. **Performance repair**: only after the circuit is legal and topology-consistent, decide whether the current topology can realistically meet spec.
4. **Minimality rule**: preserve correct substructures unless the diagnosis explicitly says the current topology has a ceiling and must be upgraded.

## Original Task Specification
**Circuit Type:** Common Gate Amplifier
**Input/Output Nodes:** subckt dut VDD VSS VIN1 VOUT1 VB1
**Topology Description:** Design an externally biased NMOS common gate amplifier with a diode-connected PMOS load to VDD.
**External Load:** A capacitor with capacitance 18 pF.
**Design Specs:** The gain should be larger than 4.64. The bandwidth should be larger than 54100000. The power should be smaller than 0.00462.

## Previous Generated Netlist
```spice
subckt dut VDD VSS VIN1 VOUT1 VB1
N0 (VOUT1 VB1 VIN1 VSS) tsmc18dN w=Wu l=Ln
P0 (VOUT1 VOUT1 VDD VDD) tsmc18dP w=Wu l=Ln
C0 (VOUT1 VSS) capacitor c=Cp
ends
```

## Failure Summary
**Failure Category:** spec_not_met_near_miss

**Metric Summary:**
gain passes (5.146 >= 4.64)
bandwidth short by 5.5% (achieved 5.111e+07, need > 5.41e+07)
power passes (0.002708 <= 0.00462)

**Diagnosis Summary:**
All key devices appear active. The miss is small and limited to bandwidth. Preserve the topology and apply only a local correction to the limiting branch.

**Additional Information:**
{
  "gain": "5.15",
  "bandwidth": "51100000",
  "power": "0.00271"
}

**Raw Error / Debug Message:**
The circuit simulates successfully but does not meet the target specs. Best results: gain=5.15, bandwidth=51100000, power=0.00271. Required spec: The gain should be larger than 4.64. The bandwidth should be larger than 54100000.0. The power should be smaller than 0.00462.

## How to Respond Based on Failure Category

### Case A: `sanity_failed`
Your job is to repair structural or connectivity problems.
Typical issues include:
- floating internal node
- MOS gate has no DC bias path
- NMOS/PMOS has no valid DC current path
- drain and source shorted together
- declared IO port unused
- output node not actively driven
- transistor role inconsistent with named topology

For this case:
- Do **not** optimize for performance first.
- Keep the intended topology from the task.
- Fix only the wrong devices / wrong nets / wrong terminal assignments.
- After each fix, mentally re-check DC current path and IO usage.

**Parser-sensitive formatting reminder for sanity repairs:**
- Use only `R0,R1,...`, `C0,C1,...`, `I0,I1,...`, `N0,N1,...`, `P0,P1,...`
- Do **not** use names like `RD`, `RG`, `RIN`, `R_D`, `Rload`
- Use only the canonical parenthesized netlist style: `R0 (net1 net2) resistor r=RK`
- Do **not** switch to shorthand SPICE syntax

### Case B: `bo_failed`
The circuit passes basic structural checks, but BO or simulation fails to produce a stable valid result, often due to fragile biasing, ambiguous DC paths, convergence-sensitive self-biasing, or invalid operating regions across most trials.
For this case:
- Do not treat this as a pure spec optimization problem.
- First check whether the topology has fragile biasing, floating-like internal control nodes, weak DC references, or ambiguous current paths.
- Prefer making the biasing and current paths more explicit.
- Do not add high-gain architecture upgrades unless the diagnosis explicitly says the topology ceiling is insufficient.

### Case C: `spec_not_met_near_miss`
The circuit is legal and topology-consistent, and it is **close** to the target.
For this case:
- Only modify a small part of the topology; no major changes are needed.
- Avoid adding or removing too many devices; prefer local adjustments within the existing topology.
- Use diagnosis + OP info to identify the limiting branch.
- Make the smallest topology edit that addresses the failing metric.
- If only one metric misses and the miss is small, avoid large architecture changes.

### Case D: `spec_not_met_major_miss`
The circuit is legal, but the gap is too large for small edits.
For this case:
- You may replace the current topology with a stronger but still task-consistent topology.
- Only do this if the diagnosis explicitly indicates a topology ceiling or clearly wrong device role.
- Keep all required task constraints: IO ports, load type, required active/passive elements, bias resources.

## Interpreting the Diagnosis Summary
The diagnosis summary is more important than raw numbers.
Use it directly.

Examples:
- "Topology ceiling insufficient for gain" -> consider cascode / higher output resistance / stronger load structure.
- "A supposed cascode/load device is in linear region" -> fix that branch before adding more devices.
- "Bandwidth already passes; gain misses slightly" -> avoid modifications that sacrifice bandwidth unless required.
- "Current ratio far from target but rout already passes" -> preserve mirror topology and change branch count / mirror structure rather than random amplifier-style edits.
- "One branch heavily loads Vout" -> repair the load branch role before adding stages.

## Rules for Using Additional Information
The operating-point data in Additional Information comes from the **best sizing result found after Bayesian Optimization**. This is not a random or initial point -- it is the best the optimizer could achieve with the given topology. Therefore:
- If a transistor is in `linear`, `off`, or `subthreshold` region, which is contrary to its intended role at this optimized point, the problem is almost certainly **structural** (wrong connection, wrong transistor role, missing bias path), not a sizing issue.
- Consider whether that transistor's drain/gate/source connections match its intended role in the topology.

If Additional Information contains operating-point observations:
- treat `region=off` as a strong sign that the device is not functioning in its intended role
- treat `region=linear` on a supposed current source / cascode / active load as a warning that output resistance is low
- large `gds` usually means the branch is resistive / low-output-resistance and may hurt gain or load the output
- low or near-zero `gm` on a supposed amplifying device suggests weak transconductance or wrong biasing
- use these observations to decide **which device role is broken**, not to do sizing

Do **not** overfit to raw numeric values.
Use them only to support topology diagnosis.

When Additional Information is provided, treat it as **branch-role evidence only**.
Do not redesign the circuit based on raw transistor numbers alone.
Use OP observations only to decide:
- which branch is likely broken
- whether a branch behaves as a current source, resistive load, or inactive branch

Do not use OP observations as a reason for broad architecture change in near-miss cases.

## Netlist Format Rules (must follow in corrected output)

### Subckt Header Preservation
The corrected netlist must keep the same `subckt` header and declared ports as the original task / previous generated netlist, unless the header itself is invalid. Do not rename ports, remove ports, reorder ports, or introduce new external ports. All corrections should be made inside the subckt body.

### Device Instance Naming
Every device instance name must follow the format **prefix + integer**, where the prefix identifies the device type:
- **NMOS transistors:** `N0`, `N1`, `N2`, ...
- **PMOS transistors:** `P0`, `P1`, `P2`, ...
- **Resistors:** `R0`, `R1`, `R2`, ...
- **Capacitors:** `C0`, `C1`, `C2`, ...
- **Current sources:** `I0`, `I1`, `I2`, ...

**Forbidden:** descriptive names like `M_cs`, `Ntail`, `Rload`, `C_out`, `PMOS_load`. Use only the short prefix+number form.

### Symbolic Parameters
- Resistors: `r=RK`; Capacitors: `c=Cp`; NMOS: `w=Wu l=Ln`; PMOS: `w=Wu l=Ln`
- Technology: `tsmc18dN` (NMOS), `tsmc18dP` (PMOS)

### MOSFET Terminal Order
**Drain, Gate, Source, Body (D G S B)** for both NMOS and PMOS.

### External Load Rule
The external load specified in **External Load** is applied by the evaluation testbench, not by the DUT subckt.
Do **not** instantiate any additional load capacitor or load resistor inside the corrected subckt netlist unless the original topology description explicitly says the load is an intrinsic circuit component.

## Edit Budget Rules
Before editing, classify the change you plan to make.

### Allowed small edits
- swap a wrong MOS terminal assignment
- reconnect a bias gate to the correct `vb*` node
- reconnect output to the correct drain/source node
- replace an invalid self-bias branch with a cleaner bias branch
- add/remove one cascode or one required active-load device only when the diagnosis explicitly identifies insufficient output resistance, a broken cascode/load role, or a topology ceiling

### Avoid these changes unless diagnosis explicitly requires it
- changing the named topology family entirely
- adding many new branches at once
- inventing extra bias networks not implied by the ports
- using fragile floating self-bias nodes when a direct `vb` bias exists

## Required Reasoning Procedure
Follow these steps internally before writing the answer:

### Step 1: Determine the failure class
Decide whether this is a structural repair, a BO robustness repair, a near-miss spec repair, or a major-miss spec repair.

### Step 2: Identify what is already correct
List the substructures that already match the task and should be preserved.

### Step 3: Identify the limiting device or branch
Use the diagnosis and additional information to identify the single most likely problematic device role / branch.

### Step 4: Choose the smallest valid fix
Prefer terminal reconnection or one local branch upgrade before broader redesign.

### Step 5: Re-check task consistency
Verify:
- all declared IO ports are used
- MOS terminal order is D G S B
- the output node is actively driven
- every MOS has a valid DC path
- the topology still matches the task description

## Output Format
Return only:
1. a short correction rationale in 3-8 bullets
2. the corrected netlist in one fenced `spice` block

Do not output sizing advice.
Do not output multiple candidate circuits.
Output one corrected topology only.

**Critical: topology changes only, no parameter tuning.**
Do not modify symbolic parameter values (e.g. do not change `Cp` to `0.95Cp`, do not change `RK` to `2*RK`).
Keep all symbolic parameters exactly as `r=RK`, `c=Cp`, `w=Wu l=Ln`.
Your correction must be a **structural/connection change** (adding, removing, or reconnecting devices), not a numeric tweak.
\end{lstlisting}

\medskip
\textbf{\texttt{"output":}}
\begin{lstlisting}[basicstyle=\ttfamily\small, breaklines=true, columns=fullflexible, keepspaces=true]
Diagnosis and Repair Reasoning:
1. The failure happened at the **spec-satisfaction stage**, not at BO or simulation. BO status is `ok`, so the topology was structurally legal and could be sized. Simulation status is also `ok`, so the circuit is electrically runnable. The problem is that **the circuit operates, but does not meet all required specs**. That means this is not a syntax/DC-connectivity failure; it is a **performance shortfall** in an otherwise valid architecture.

2. The reported evidence says the shortfall is **narrow and specific**:
- Required gain > 4.64, achieved gain = **5.146** -> gain is already good enough.
- Required bandwidth > 54.1 MHz, achieved bandwidth = **51.1 MHz** -> bandwidth is the only missed metric.
- Required power < 0.005, achieved power = **0.003** -> power is comfortably fine.

So the circuit is not fundamentally the wrong amplifier family. It already behaves like a usable gain stage with acceptable gain and low power. The issue is instead that the realized structure is **slightly too slow** for the required capacitive load. Since gain is fine and power is low, the diagnosis should focus on whether the topology includes any **unnecessary loading element or extra pole** that reduces output bandwidth.

3. Looking at the wrong netlist, the active core is:
- `N0 (VOUT1 VB1 VIN1 VSS)` -> NMOS common-gate device
- `P0 (VOUT1 VOUT1 VDD VDD)` -> PMOS diode-connected load

That part is actually consistent with the intended externally biased NMOS common-gate amplifier with diode-connected PMOS load. The suspicious part is the added capacitor:
- `C0 (VOUT1 VSS) capacitor c=Cp`

In the original task, the external load is indeed a capacitor, so including a load capacitor is not conceptually wrong in a standalone design description. But here the simulation evidence is very telling: gain and power are already acceptable, while bandwidth is only modestly below target. In a common-gate stage with diode-connected PMOS load, the dominant bandwidth limitation is usually the **output pole at VOUT1**. Adding an explicit capacitor directly from `VOUT1` to `VSS` makes that pole heavier. Since the topology itself is already correct, this extra capacitive burden is the most direct explanation for why the circuit misses only bandwidth.

4. The repair direction therefore is **not to change the amplifier architecture**, but to **remove the unnecessary explicit output capacitor from the submitted netlist**. The common-gate NMOS plus diode-connected PMOS load is already the right gain path. The feedback does not suggest missing biasing, wrong output node, or wrong transistor roles. Instead, it suggests that the implemented circuit is the intended amplifier **plus extra output loading**, which slows it down enough to fail the bandwidth target.

5. Structurally, the corrected circuit should keep:
- the NMOS as the **common-gate transconductor**: drain at `VOUT1`, gate at `VB1`, source at `VIN1`
- the PMOS as the **diode-connected pull-up load**: drain/gate at `VOUT1`, source at `VDD`

What must change is that `VOUT1` should remain only the intrinsic output node of that active NMOS/PMOS pair, rather than also being burdened by an added explicit capacitor inside the subcircuit. In other words, the wrong circuit does not need to be reorganized into a different topology; it needs to be **trimmed back to the intended two-transistor common-gate stage** so that the output pole is not artificially worsened. The gain path stays the same, the load role stays the same, and the bias role stays the same; only the extra bandwidth-limiting element should be removed.

6. The reference netlist confirms exactly that diagnosis. It contains only the NMOS common-gate device and the diode-connected PMOS load, with no explicit capacitor:
- NMOS: `(VOUT1 VB1 VIN1 VSS)`
- PMOS: `(VOUT1 VOUT1 VDD VDD)`

So the inferred repair direction matches the true solution direction: the wrong netlist was already architecturally correct in its transistor structure, but it included an extra output capacitor that pushed bandwidth below spec.

Corrected Complete Netlist:
```spice
    subckt dut VDD VSS VIN1 VOUT1 VB1
    N0 (VOUT1 VB1 VIN1 VSS) tsmc18dN w=Wu l=Ln
    P0 (VOUT1 VOUT1 VDD VDD) tsmc18dP w=Wu l=Ln
    ends dut
```
\end{lstlisting}

\textbf{\texttt{\}}}
\end{tcolorbox}
    
\end{center}

\section{Implementation Details and Additional Experimental Results}
\subsection{Implementation Details}
\label{app:imple_details}

The detailed recipes for three-stage SFT and simulation feedback-augmented RFT are reported in Tab.~\ref{tab:sft_hyperparams} and Tab.~\ref{tab:rft_hyperparams}, respectively. We also provide one training example for each stage of our three-stage SFT curriculum. In addition, we include the prompt templates used for single-pass generation and two-pass generation with simulation feedback.

\begin{table}[ht]
\caption{\textbf{Detailed three-stage SFT recipes.}}
\label{tab:sft_hyperparams}
\centering{
\resizebox{1\linewidth}{!}{
\begin{tabular}{lccc}
\toprule
 & \textbf{Knowledge Acquisition} & \textbf{Component Prediction} & \textbf{Simulation Feedback-Guided Correction} \\
\midrule
Epochs & 1 & 1 & 1 \\
Total Effective Batch Size & 64 & 64 & 64 \\
Warm-up Ratio & 0.00 & 0.00 & 0.00 \\
Learning Rate & 1e-5 & 1e-5 & 1e-5 \\
GPU Hardware & \multicolumn{3}{c}{8x NVIDIA RTX PRO 6000} \\
\bottomrule
\end{tabular}}}
\end{table}

\begin{table}[ht]
\caption{\textbf{Detailed RFT recipes.}}
\label{tab:rft_hyperparams}
\centering{
\resizebox{0.6\linewidth}{!}{
\begin{tabular}{lr}
\toprule
Optimizer & AdamW\\
Total Batch Size & 640\\
Mini Batch Size & 320 \\
Learning Rate & 1e-6\\
LR Schedule & Constant \\
Weight Decay & 0 \\
Warm-up Ratio & 0 \\
Number of Training Steps & 100 \\
Number of Gradient Steps & 1600 \\
Max Prompt Length & 5000 \\
Max Rollout Length & 4096 \\
Number of Rollouts Per Prompt & 8\\
Rollout Sampling Temperature & 0.6 \\
Rollout Sampling Top-p & 0.95 \\
GPU Hardware & 8x  NVIDIA RTX PRO 6000  \\
\bottomrule
\end{tabular}}}
\end{table}

\begin{tcolorbox}[breakable, title={Single-Pass Generation Template}]
\begin{lstlisting}[basicstyle=\ttfamily\small, breaklines=true, columns=fullflexible, keepspaces=true]
# Circuit Topology Design Task

## Objective
Design a circuit topology based on the given circuit type and specifications. Ensure the designed circuit is functionally correct and meets all design requirements.

## Design Examples
The examples below illustrate the simplest single-stage topologies (CD, CG) with basic loads. Real design tasks may require more complex structures, like cascode stages, regulated cascode, multi-stage amplifiers, differential pairs, etc. They would involve more transistors and internal nodes. The design methodology (Step 1-4) and all rules still apply; treat these examples as the building blocks that more complex topologies are composed from.

### Example 1: Resistor-Loaded Common Drain Amplifier

**Circuit Type:**  Common Drain Amplifier 
**Input/Output Nodes:**  subckt DUT VDD VSS vin vout
**Topology Description:** Design a resistor-loaded common drain (source follower) amplifier. The topology uses an NMOS source follower transistor with a resistor as the load. 
**External load:** 'C' pF. *(provided by the testbench at simulation time -- do NOT include this capacitor inside the subckt.)*
**Design Specs:**  Gain, Bandwidth

**Step 1: Analyze Nodes and Circuit Type**
- Available nodes: VDD, VSS, vin, vout
- Bias voltage node (vb/vbias): Not available
- Current reference node (ib/ibp/ibn): Not available
- Topology description indicates: resistor load, single NMOS source follower
- Topology selection: Resistor-loaded common drain amplifier (only option given no bias nodes)

**Step 2: Identify Components**
- M0 (NMOS): Source follower transistor (gate = vin, drain = VDD, source = vout)
- R0: Load resistor (vout to VSS) -- **part of the topology** (without it the source follower has no DC current path), so it MUST appear in the netlist.
- *(External load capacitor C is provided by the testbench at simulation time -- do NOT include it in the netlist.)*

**Step 3: Define Connections**
M0 (NMOS Source Follower):
Drain -> VDD
Gate -> vin (input signal)
Source -> vout (output node)
Body -> VSS

R0 (Load Resistor):
Terminal 1 -> vout
Terminal 2 -> VSS

**Step 4: Output Netlist**
subckt DUT VDD VSS vin vout
    N0 (VDD vin vout VSS) tsmc18dN w=Wu l=Ln 
    R0 (vout VSS) resistor r=RK
ends DUT

### Example 2: Common Gate Amplifier with PMOS Active Load

**Circuit Type:** Common Gate Amplifier  
**Input/Output Nodes:** subckt DUT VDD VSS vin vb vout
**Topology Description:** Design a common gate amplifier with active load. 
**External load:** 'C' pF. *(provided by the testbench at simulation time -- do NOT include this capacitor inside the subckt.)*
**Design Specs:** Gain, Bandwidth

**Step 1: Analyze Nodes and Circuit Type**
- Available nodes: VDD, VSS, vin, vb, vout
- Bias voltage node (vb): Available -> vb can serve multiple roles depending on the topology: bias a common-gate transistor's gate, bias a cascode transistor's gate, or bias an active load / tail current source. The specific role is determined by the topology description. Here, the description specifies CG -> vb biases the CG transistor's gate.
- Current reference node (ib/ibp/ibn): Not available
- Topology description indicates: NMOS common-gate, PMOS diode-connected load
- Topology selection: Common gate amplifier with PMOS active load
- **Common Gate key feature:** The NMOS **gate is biased to a fixed voltage** (vb), the **input is applied to the source** terminal, and the **output is taken from the drain** terminal. Unlike Common Drain, the drain does NOT connect to VDD -- it connects to the output node.

**Step 2: Identify Components**
- N0 (NMOS): Common-gate transistor (gate = vb, source = vin, drain = vout)
- P0 (PMOS): Diode-connected active load (gate = drain = vout, source = VDD)
- *(External load capacitor C is provided by the testbench at simulation time -- do NOT include it in the netlist.)*

**Step 3: Define Connections**
N0 (NMOS Common-Gate):
Drain -> vout (output node -- NOT VDD)
Gate -> vb (fixed DC bias -- this is what makes it "common gate")
Source -> vin (input signal applied at source terminal)
Body -> VSS

P0 (PMOS Diode-Connected Load):
Drain -> vout (same node as N0 drain; provides DC path from VDD to vout)
Gate -> vout (gate shorted to drain = diode-connected)
Source -> VDD
Body -> VDD

**DC current path check:**
- N0 (NMOS): VDD -> P0(source->drain) -> vout = N0.drain -- N0.source = vin -> (external DC bias on vin provides path to VSS)
- P0 (PMOS): VDD -> P0.source -- P0.drain = vout -> through N0 to VSS

**Step 4: Output Netlist**
subckt DUT VDD VSS vin vb vout
    N0 (vout vb vin VSS) tsmc18dN w=Wu l=Ln 
    P0 (vout vout VDD VDD) tsmc18dP w=Wu l=Ln 
ends DUT

### Example 3: Simple NMOS Current Mirror

**Circuit Type:** Current Mirror  
**Input/Output Nodes:** subckt DUT VDD VSS IOUT1 Ibp10u  
**External Load:** No external load. Do not add any resistor or capacitor in between the output node and ground (VSS), or any current source, unless explicitly required.  
**Topology Description:** Design a simple NMOS current mirror that converts an input reference current into an output current.

**Step 1: Analyze Nodes and Circuit Type**
- Available nodes: VDD, VSS, IOUT1, Ibp10u
- Bias voltage node (vb*): Not available
- Current reference node (ib* / ibp* / ibn*): Ibp10u is available
- Topology description indicates: simple NMOS current mirror
- Key current mirror interpretation:
  - Ibp10u is the reference current port, not an ordinary internal node
  - IOUT1 is the output current port, not a voltage-output node like vout
  - Since the description says simple current mirror, the topology should contain:
    - one diode-connected NMOS reference transistor
    - one NMOS mirror transistor
  - No extra resistor, capacitor, or current source should be invented

**Step 2: Identify Components**
- N0 (NMOS): diode-connected reference transistor
- N1 (NMOS): mirror output transistor

**Step 3: Define Connections**
N0 (Reference NMOS):
Drain -> Ibp10u  
Gate -> Ibp10u  
Source -> VSS  
Body -> VSS

N1 (Mirror NMOS):
Drain -> IOUT1  
Gate -> Ibp10u  
Source -> VSS  
Body -> VSS

**Current mirror self-check:**
- N0 is diode-connected because drain = gate = Ibp10u
- N1 shares the same gate bias as N0 through node Ibp10u
- N1 drain is the output current node IOUT1
- Both NMOS sources and bodies connect to VSS
- No extra load branch is added

**Step 4: Output Netlist**
subckt DUT VDD VSS IOUT1 Ibp10u
    N0 (Ibp10u Ibp10u VSS VSS) tsmc18dN w=Wu l=Ln
    N1 (IOUT1 Ibp10u VSS VSS) tsmc18dN w=Wu l=Ln
ends DUT

## Design Rules

### Rule 1: Use Symbolic Parameters (No Numerical Values)
Use these Symbolic Parameters for ALL components:
    - Resistors: r=RK (not actual values) resistor
    - Capacitors: c=Cp (not actual values) capacitor
    - NMOS transistors: w=Wu l=Ln (not actual values)
    - PMOS transistors: w=Wu l=Ln (not actual values)
    - Technology: tsmc18dN (NMOS), tsmc18dP (PMOS)

**Technology:** TSMC 180nm process(tsmc18dN and tsmc18dP)

### Rule 1B: Never Write Numeric Device or Load Values
Even if the task description mentions concrete values such as 5MOhm, 16 pF, 7 pF, or 0.1 pF, do not write numeric values into the returned netlist.

Still use only symbolic parameters:
- resistor -> r=RK
- capacitor -> c=Cp
- MOSFET sizes -> w=Wu l=Ln

Do not output values like r=5e6, c=16p, or c=0.1p.

### Rule 2: MOSFET Terminal Order
The terminal order is **Drain, Gate, Source, Body (D G S B)** for both NMOS and PMOS.

### Rule 3: Passive Component Polarity
Resistors and capacitors have no polarity requirement.  
R0 (net1 net2) is equivalent to R0 (net2 net1).

### Rule 4: Node Naming Convention
VDD: Supply Voltage
VSS: Ground
vout: Output Voltage
Voltage nodes start with vb stands for biased voltage
Current node: 
1. ibp. current flowing out. You could consider it as a current source providing current from VDD to this ibp* node
2. ibn. current flowing in. You could consider it as a current source providing current this ibn* node to VSS
3. ib/ibias. Unspecified direction
If only vin shown in Input/output node, it stands for input voltage
vin, vip stands for input voltage (negative and positive)

**Current source connection examples:**
- For ibp* nodes: I0 (VDD ibp10u) isource dc=IA
- For ibn* nodes: I0 (ibn5u VSS) isource dc=IA

### Port-to-Topology Reasoning Guide

When analyzing the subckt IO ports, use this reasoning framework:

**Step A: Identify available resources**
- Count how many vb* ports -> each provides a free DC bias voltage, which can bias a MOSFET gate (for cascode, common-gate, current source, or active load)
- Count how many ibp*/ibn* ports -> each provides a reference current, typically used to build a current mirror for biasing
- If neither vb nor ib ports exist -> the design uses internally generated biasing, such as resistive loads or diode-connected MOSFETs.

**Step B: Match resources to topology needs**
- A single-stage CS/CG/CD amplifier with active load typically needs 1 bias resource (either vb for the load gate, or ib for a mirror)
- A cascode amplifier typically needs 1-2 extra vb ports for cascode gate biasing
- A differential pair with tail current source needs 1 bias resource for the tail (vb or ib)
- Multiple vb ports (vb1, vb2) often indicate a cascode or regulated cascode topology

**Step C: Handle resource mismatches**
Sometimes the available ports do not directly match what the topology needs. Common conversions:

- Need a bias current, but only have vb:
Use vb to directly drive a current-source MOSFET's gate. No current mirror is needed -- vb sets the gate voltage and thus the drain current. 
Example: N_tail (source_node vb VSS VSS) creates a current source biased by vb. 

- Need a bias voltage, but only have ibp/ibn:
Use a diode-connected MOSFET to convert the reference current into a gate voltage. The diode-connected device's gate-drain node acts as the bias voltage, which can then be shared to other gates. 
Example: N_ref (ibp10u ibp10u VSS VSS) converts ibp current into a voltage at the ibp10u node, which mirrors to other NMOS gates.

- Have neither vb nor ib:
Use passive loads (resistors) or self-biased structures (diode-connected MOSFETs wired within the signal path, e.g., diode-connected PMOS load on a CS amplifier).

**Step D: Read the topology description carefully**
- The topology description specifies which devices to use and how they connect. Use Steps A-B to understand the resources, then follow the description for the specific implementation.

### Rule 5: Device Instance Naming
Every device instance name must follow the format **prefix + integer**, where the prefix identifies the device type:
- NMOS transistors: N0, N1, N2, ...
- PMOS transistors: P0, P1, P2, ...
- Resistors: R0, R1, R2, ...
- Capacitors: C0, C1, C2, ...
- Current sources: I0, I1, I2, ...

**Forbidden:** descriptive names like M_cs, Ntail, Rload, C_out, PMOS_load. Use only the short prefix+number form.

### Rule 5B: Netlist Syntax Only
Use only the parenthesized netlist syntax shown in the examples.

Required forms:
- NMOS: N0 (drain gate source body) tsmc18dN w=Wu l=Ln
- PMOS: P0 (drain gate source body) tsmc18dP w=Wu l=Ln
- Resistor: R0 (net1 net2) resistor r=RK
- Capacitor: C0 (net1 net2) capacitor c=Cp
- Current source: I0 (net1 net2) isource dc=IA

Do not use alternative shorthand SPICE forms such as:
- RD VDD vout RK
- M0 VDD Vin Vout VSS ...
- mixed parenthesized / non-parenthesized syntax

Always use the parenthesized style only.

### Rule 6: Net Name Format
Net names (including names for internal nodes) must contain only letters, digits, and underscores.
- Allowed: net1, source_node, nodeA, drain_2
- Forbidden: parentheses (), brackets [], slashes /\, spaces, hyphens -, dots ., special characters like Ohm, micro
- Bad examples: (VIN), net-1, node.A, 5MOhm

### Rule 7: Use ALL Declared IO Ports
Every port declared in the subckt line must appear in at least one device connection inside the netlist. If the subckt declares subckt DUT VDD VSS vin vb vout, then vin, vb, and vout must all be used as pin connections in at least one device. Do not declare ports you do not use.

### Rule 8: Common Design Problems
1. Tie NMOS bodies to VSS and PMOS bodies to VDD unless the task explicitly specifies otherwise.
2. Do not treat ibp*/ibn* nodes as ordinary internal nets; use them only as external current-source ports or current-reference ports.
3. Ensure every critical node (especially vout and any bias node) has a DC path to VDD or VSS; avoid floating nodes.
4. Never rely on a capacitor alone to provide DC bias; capacitors do not create a DC conduction path.
5. Place PMOS devices on pull-up paths to VDD and NMOS devices on pull-down paths to VSS to avoid invalid headroom and body-diode conduction.

### Rule 8B: Do Not Add Output Loads or Helper Branches

Your netlist must NOT contain ANY of the following:

1. A capacitor connecting an output port (e.g. IOUT*, vout, vop, von) to ground (VSS).
   Examples of forbidden lines:
       C0 (vout VSS) capacitor c=Cp
       C0 (IOUT1 VSS) capacitor c=Cp

   Note: a load resistor between the output and VSS is allowed only when it is part of the declared topology (e.g. the load resistor of a resistor-loaded common-drain amplifier as in Example 1). External load capacitors are always provided by the testbench, not by the netlist.

2. Any voltage source or current source device.
   Examples of forbidden lines:
       I0 (VDD ibp10u) isource dc=IA
       V0 (vb VSS) vsource dc=Vbias

**Rationale:** All supplies, bias voltages, bias currents, and load capacitors are provided by the external testbench at simulation time -- the netlist itself should contain only the topology devices (MOSFETs, plus any resistors/capacitors that are explicitly part of the declared topology).

### Rule 9: Output Node Must Have Active Drive
The voltage output node (vout / VOUT) must connect directly to at least one MOSFET's drain or source terminal or have a path to drain/source. An output driven only through passive components (resistors, capacitors) cannot provide signal amplification or buffering. 

### Rule 10: Every MOSFET Must Have a Valid DC Current Path
For each MOSFET in the circuit:
1. Drain and source must connect to different nets. A MOSFET with drain = source forms a MOS capacitor, not an active transistor.
2. NMOS: drain must have a DC path to VDD (through load resistor, PMOS, or other devices), and source must have a DC path to VSS. The complete current path is: VDD -> load -> drain -- source -> VSS.
3. PMOS: source must have a DC path to VDD, and drain must have a DC path to VSS. The complete current path is: VDD -> source -- drain -> load -> VSS.
4. Gate must have a DC bias path. The gate must connect to a signal input (vin), bias port (vb), diode-connected device, or resistive divider -- not be left floating or connected only through a capacitor.

### Rule 11: No Floating Internal Nodes
Every internal net (not declared as an IO port) must connect to at least two device terminals. A net that appears only once is a dangling wire. Additionally:
- A net connected only to MOSFET gates (and nothing else) has no DC bias -- it is floating.
- A net connected only to MOSFET gates and capacitors is also DC-floating, because capacitors block DC.

### Rule 12: MOSFET Drain Should Usually Not Connect Directly to VDD or VSS (Except Common Drain)

In most amplifier topologies, a MOSFET's drain connects to an output node or an internal circuit node -- not directly to VDD or VSS.

The main exception is Common Drain (source follower), where the drain is intentionally tied to the supply (VDD for NMOS, VSS for PMOS).

**Self-check:** For each MOSFET, if its drain is wired directly to VDD or VSS, verify that this is intentional (this usually indicates a source follower). If it is intended to be a gain stage, cascode, current source, or active load, the drain should connect to a circuit node instead.

Note: In complex topologies (cascode, multi-stage), many MOSFETs will have drains connecting to internal nodes -- this is normal and expected. Not every drain connects to vout; only the final output stage's relevant device does.

## Your Task

**Circuit Type:** [NAME]
**Description** [DES]
**Input/Output Nodes:** [HEAD]
**External Load** [LOADER]
**Design Specs:** [SPEC]

## Output Format
1. Provide a brief analysis of the task (node identification and topology construction)
2. Return the complete netlist within code blocks, starting with subckt and ending with ends
\end{lstlisting}
\end{tcolorbox}

\label{app:single_pass_inference_template}

\begin{tcolorbox}[breakable, title={Two-pass Generation with Simulation Feedback Template}]
\begin{lstlisting}[basicstyle=\ttfamily\small, breaklines=true, columns=fullflexible, keepspaces=true]
# Feedback Correction Task

## Objective
You are correcting a previously generated circuit netlist.
Your goal is **not** to redesign from scratch by default. First determine the failure type, then apply the **smallest topology edit** that can realistically fix the failure.

You will be given:
- the original task specification
- the previous incorrect/generated netlist
- a structured failure category
- a concise metric summary
- a diagnosis summary
- optional operating-point observations
- optional BO/debug summary

## Priority Order
Follow this decision order strictly:
1. **Legality**: fix floating nodes, unused ports, impossible DC paths, wrong terminal roles, broken source/drain connections.
2. **Topology consistency**: make sure the circuit actually matches the named topology and topology description.
3. **Performance repair**: only after the circuit is legal and topology-consistent, decide whether the current topology can realistically meet spec.
4. **Minimality rule**: preserve correct substructures unless the diagnosis explicitly says the current topology has a ceiling and must be upgraded.

## Original Task Specification
**Circuit Type:** [NAME]
**Input/Output Nodes:** [HEAD]
**Topology Description:** [DES]
**External Load:** [LOADER]
**Design Specs:** [SPEC]

## Previous Generated Netlist
```spice
[BAD_GENERATION]
```

## Failure Summary
**Failure Category:** [FAILURE_CATEGORY]

**Metric Summary:**
[METRIC_SUMMARY]

**Diagnosis Summary:**
[DIAGNOSIS]

**Additional Information:**
[INFO]

**Raw Error / Debug Message:**
[ERROR_MESSAGE]

## How to Respond Based on Failure Category

### Case A: `sanity_failed`
Your job is to repair structural or connectivity problems.
Typical issues include:
- floating internal node
- MOS gate has no DC bias path
- NMOS/PMOS has no valid DC current path
- drain and source shorted together
- declared IO port unused
- output node not actively driven
- transistor role inconsistent with named topology

For this case:
- Do **not** optimize for performance first.
- Keep the intended topology from the task.
- Fix only the wrong devices / wrong nets / wrong terminal assignments.
- After each fix, mentally re-check DC current path and IO usage.

**Parser-sensitive formatting reminder for sanity repairs:**
- Use only `R0,R1,...`, `C0,C1,...`, `I0,I1,...`, `N0,N1,...`, `P0,P1,...`
- Do **not** use names like `RD`, `RG`, `RIN`, `R_D`, `Rload`
- Use only the canonical parenthesized netlist style: `R0 (net1 net2) resistor r=RK`
- Do **not** switch to shorthand SPICE syntax

### Case B: `bo_failed`
The circuit passes basic structural checks, but BO or simulation fails to produce a stable valid result, often due to fragile biasing, ambiguous DC paths, convergence-sensitive self-biasing, or invalid operating regions across most trials.
For this case:
- Do not treat this as a pure spec optimization problem.
- First check whether the topology has fragile biasing, floating-like internal control nodes, weak DC references, or ambiguous current paths.
- Prefer making the biasing and current paths more explicit.
- Do not add high-gain architecture upgrades unless the diagnosis explicitly says the topology ceiling is insufficient.

### Case C: `spec_not_met_near_miss`
The circuit is legal and topology-consistent, and it is **close** to the target.
For this case:
- Only modify a small part of the topology; no major changes are needed.
- Avoid adding or removing too many devices; prefer local adjustments within the existing topology.
- Use diagnosis + OP info to identify the limiting branch.
- Make the smallest topology edit that addresses the failing metric.
- If only one metric misses and the miss is small, avoid large architecture changes.

### Case D: `spec_not_met_major_miss`
The circuit is legal, but the gap is too large for small edits.
For this case:
- You may replace the current topology with a stronger but still task-consistent topology.
- Only do this if the diagnosis explicitly indicates a topology ceiling or clearly wrong device role.
- Keep all required task constraints: IO ports, load type, required active/passive elements, bias resources.

## Interpreting the Diagnosis Summary
The diagnosis summary is more important than raw numbers.
Use it directly.

Examples:
- "Topology ceiling insufficient for gain" -> consider cascode / higher output resistance / stronger load structure.
- "A supposed cascode/load device is in linear region" -> fix that branch before adding more devices.
- "Bandwidth already passes; gain misses slightly" -> avoid modifications that sacrifice bandwidth unless required.
- "Current ratio far from target but rout already passes" -> preserve mirror topology and change branch count / mirror structure rather than random amplifier-style edits.
- "One branch heavily loads Vout" -> repair the load branch role before adding stages.

## Rules for Using Additional Information
The operating-point data in Additional Information comes from the **best sizing result found after Bayesian Optimization**. This is not a random or initial point -- it is the best the optimizer could achieve with the given topology. Therefore:
- If a device is in `linear`, `off`, or `subthreshold` at this optimized point, the problem is almost certainly **structural** (wrong connection, wrong device role, missing bias path), not a sizing issue.
- Consider whether that device's drain/gate/source connections match its intended role in the topology.

If Additional Information contains operating-point observations:
- treat `region=off` as a strong sign that the device is not functioning in its intended role
- treat `region=linear` on a supposed current source / cascode / active load as a warning that output resistance is low
- large `gds` usually means the branch is resistive / low-output-resistance and may hurt gain or load the output
- low or near-zero `gm` on a supposed amplifying device suggests weak transconductance or wrong biasing
- use these observations to decide **which device role is broken**, not to do sizing

Do **not** overfit to raw numeric values.
Use them only to support topology diagnosis.

When Additional Information is provided, treat it as **branch-role evidence only**.
Do not redesign the circuit based on raw transistor numbers alone.
Use OP observations only to decide:
- which branch is likely broken
- whether a branch behaves as a current source, resistive load, or inactive branch

Do not use OP observations as a reason for broad architecture change in near-miss cases.

## Netlist Format Rules (must follow in corrected output)

### Subckt Header Preservation
The corrected netlist must keep the same `subckt` header and declared ports as the original task / previous generated netlist, unless the header itself is invalid. Do not rename ports, remove ports, reorder ports, or introduce new external ports. All corrections should be made inside the subckt body.

### Device Instance Naming
Every device instance name must follow the format **prefix + integer**, where the prefix identifies the device type:
- **NMOS transistors:** `N0`, `N1`, `N2`, ...
- **PMOS transistors:** `P0`, `P1`, `P2`, ...
- **Resistors:** `R0`, `R1`, `R2`, ...
- **Capacitors:** `C0`, `C1`, `C2`, ...
- **Current sources:** `I0`, `I1`, `I2`, ...

**Forbidden:** descriptive names like `M_cs`, `Ntail`, `Rload`, `C_out`, `PMOS_load`. Use only the short prefix+number form.

### Symbolic Parameters
- Resistors: `r=RK`; Capacitors: `c=Cp`; NMOS: `w=Wu l=Ln`; PMOS: `w=Wu l=Ln`
- Technology: `tsmc18dN` (NMOS), `tsmc18dP` (PMOS)

### MOSFET Terminal Order
**Drain, Gate, Source, Body (D G S B)** for both NMOS and PMOS.

### External Load Rule
The external load specified in **External Load** is applied by the evaluation testbench, not by the DUT subckt.
Do **not** instantiate any additional load capacitor or load resistor inside the corrected subckt netlist unless the original topology description explicitly says the load is an intrinsic circuit component.

## Edit Budget Rules
Before editing, classify the change you plan to make:

### Allowed small edits
- swap a wrong MOS terminal assignment
- reconnect a bias gate to the correct `vb*` node
- reconnect output to the correct drain/source node
- replace an invalid self-bias branch with a cleaner bias branch
- add/remove one cascode or one required active-load device only when the diagnosis explicitly identifies insufficient output resistance, a broken cascode/load role, or a topology ceiling

### Avoid unless diagnosis explicitly requires it
- changing the named topology family entirely
- adding many new branches at once
- inventing extra bias networks not implied by the ports
- using fragile floating self-bias nodes when a direct `vb` bias exists

## Required Reasoning Procedure
Follow these steps internally before writing the answer:

### Step 1: Determine the failure class
Decide whether this is a structural repair, a BO robustness repair, a near-miss spec repair, or a major-miss spec repair.

### Step 2: Identify what is already correct
List the substructures that already match the task and should be preserved.

### Step 3: Identify the limiting device or branch
Use the diagnosis and additional information to identify the single most likely problematic device role / branch.

### Step 4: Choose the smallest valid fix
Prefer terminal reconnection or one local branch upgrade before broader redesign.

### Step 5: Re-check task consistency
Verify:
- all declared IO ports are used
- MOS terminal order is D G S B
- the output node is actively driven
- every MOS has a valid DC path
- the topology still matches the task description

## Output Format
Return only:
1. a short correction rationale in 3-8 bullets
2. the corrected netlist in one fenced `spice` block

Do not output sizing advice.
Do not output multiple candidate circuits.
Output one corrected topology only.

**Critical: topology changes only, no parameter tuning.**
Do not modify symbolic parameter values (e.g. do not change `Cp` to `0.95Cp`, do not change `RK` to `2*RK`).
Keep all symbolic parameters exactly as `r=RK`, `c=Cp`, `w=Wu l=Ln`.
Your correction must be a **structural/connection change** (adding, removing, or reconnecting devices), not a numeric tweak.
\end{lstlisting}
\end{tcolorbox}

\label{app:two_pass_inference_template}

\subsection{Additional Case Study}
\label{app:case_study}
We provide further details for the four qualitative examples in Fig.~\ref{fig:qualitative_example}. 
These examples cover two failure categories in our validation pipeline: circuit topology check failures and unmet-specification failures. 
For each case, we describe the design task, the failure mode of the initial generation, and the concrete revision made after receiving simulation-grounded validation feedback.

\paragraph{Case A --- Topology-Check Failure: NMOS Cascode Current Mirror.}
This case corresponds to Fig.~\ref{fig:qualitative_example}(a), where the model is asked to design an NMOS cascode current mirror with ports \texttt{VDD/VSS/Iout/Vb/Iref} and a target current ratio of $10.5$. The initial generation contains a structurally invalid cascode branch:
\begin{lstlisting}[basicstyle=\ttfamily\small, breaklines=true, columns=fullflexible, keepspaces=true]
M1 (Iref, Iref, VSS,  VSS)
M2 (Iout, Iref, VSS,  VSS)
M3 (Iout, Vb,   Iout, VSS)
\end{lstlisting}
Here, \texttt{M3} has both its drain and source connected to \texttt{Iout}, so it does not form a valid current-carrying branch and fails the topology check before sizing or simulation. Given the validation feedback, the model identifies that \texttt{M3} is intended to serve as the cascode device above the mirror transistor \texttt{M2}, and introduces an internal node to separate the drain of \texttt{M2} from the source of \texttt{M3}:
\begin{lstlisting}[basicstyle=\ttfamily\small, breaklines=true, columns=fullflexible, keepspaces=true]
M1 (Iref, Iref, VSS,  VSS)
M2 (net1, Iref, VSS,  VSS)
M3 (Iout, Vb,   net1, VSS)
\end{lstlisting}
This edit restores the intended stacked cascode structure while preserving the device count and bias connections. After BO-based sizing and SPICE simulation, the revised design satisfies all target specifications, including output resistance, output-voltage compliance, and the required current ratio. This example shows that \method{} can use topology-check feedback to recover a latent intended structure by making a minimal node-level correction, rather than redesigning the circuit from scratch.

\paragraph{Case B --- Topology-Check Failure: Common Gate Amplifier.}
This case corresponds to Fig.~\ref{fig:qualitative_example}(b), where the model is asked to design a common-gate amplifier with a PMOS active load and ports \texttt{VDD/VSS/Vin/Vout/Vb}. The initial generation has the correct high-level structure, but leaves the declared input port \texttt{Vin} unused:
\begin{lstlisting}[basicstyle=\ttfamily\small, breaklines=true, columns=fullflexible, keepspaces=true]
M1 (Vout, Vb,   net4, VSS)
M2 (Vout, Vout, VDD,  VDD)
R1 (net4, VSS)
\end{lstlisting}
In a common gate amplifier, the input signal should be applied at the source of the NMOS input device. However, \texttt{M1}'s source is connected to the internal node \texttt{net4}, while the declared input port \texttt{Vin} is disconnected from the circuit. As a result, the design fails the topology check before sizing or simulation.

Given the validation feedback, the model identifies this as an interface-level connectivity error rather than a wrong topology. It reroutes the source of \texttt{M1} from \texttt{net4} to \texttt{Vin}, and updates the source-bias resistor consistently:
\begin{lstlisting}[basicstyle=\ttfamily\small, breaklines=true, columns=fullflexible, keepspaces=true]
M1 (Vout, Vb,  Vin, VSS)
M2 (Vout, Vout, VDD, VDD)
R1 (Vin, VSS)
\end{lstlisting}
This local edit reconnects the declared input port to the common-gate signal path while preserving the PMOS active load and the source-bias structure. After BO-based sizing and SPICE simulation, the revised design satisfies all target specifications. This example shows that \method{} can use topology-check feedback to repair interface-level defects through minimal node-level edits without unnecessarily changing the circuit architecture.

\paragraph{Case C --- Unmet-Specification Failure: NMOS Cascode Current Mirror.}
This case corresponds to Fig.~\ref{fig:qualitative_example}(c), where the model is asked to design an NMOS cascode current mirror with ports \texttt{VDD/VSS/Iout/Iref} and a target current ratio of $9.0$. The initial generation is structurally valid and implements a self-biased cascode mirror:
\begin{lstlisting}[basicstyle=\ttfamily\small, breaklines=true, columns=fullflexible, keepspaces=true]
M1 (Iref, Iref, VSS,  VSS)
M2 (net1, Iref, VSS,  VSS)
M3 (Iout, Iref, net1, VSS)
\end{lstlisting}
After BO-based sizing and SPICE simulation, the circuit satisfies the output resistance and output-voltage compliance constraints, but its current ratio reaches only $7.68$, below the target value of $9.0$. This indicates that the failure is not caused by an invalid topology or insufficient cascoding, but by inadequate output-branch scaling.

Given the simulation feedback, the model makes a targeted metric-driven revision. It preserves the existing cascode mirror structure and adds one additional output-side transistor in parallel with \texttt{M3}, reusing the same drain, gate, and source nodes:
\begin{lstlisting}[basicstyle=\ttfamily\small, breaklines=true, columns=fullflexible, keepspaces=true]
M1 (Iref, Iref, VSS,  VSS)
M2 (net1, Iref, VSS,  VSS)
M3 (Iout, Iref, net1, VSS)
M4 (Iout, Iref, net1, VSS)
\end{lstlisting}
This local edit increases the effective output-side drive strength without introducing new ports, bias nodes, or a different mirror architecture. After BO-based sizing on the revised topology, the current ratio improves from $7.68$ to $9.00$, while the output resistance and voltage-compliance metrics remain within specification. This example shows that \method{} can use simulation feedback to identify the specific failing metric and apply a minimal topology-level modification that repairs the failure while preserving already valid circuit behavior.

\paragraph{Case D --- Unmet-Specification Failure: Common Gate Amplifier.}
This case corresponds to Fig.~\ref{fig:qualitative_example}(d), where the model is asked to design a common-gate amplifier with ports \texttt{VDD/VSS/Vin/Vout/Vb1/Vb2}. The initial generation is structurally valid and passes the topology check, but its PMOS load network uses an internally derived bias node:
\begin{lstlisting}[basicstyle=\ttfamily\small, breaklines=true, columns=fullflexible, keepspaces=true]
M1 (net1, Vb1,  VDD,  VDD)
M2 (Vout, net1, VDD,  VDD)
M3 (Vout, Vb2,  Vin,  VSS)
R1 (net1, Vout)
\end{lstlisting}
Here, \texttt{M3} forms the common gate input device, while \texttt{M2} serves as the PMOS load with its gate driven by the internal node \texttt{net1}. Although the circuit is legal and can be sized, the achieved gain is only $22.9\,\mathrm{dB}$, below the required $23.4\,\mathrm{dB}$. In contrast, bandwidth ($1.40\,\mathrm{MHz}$) and power ($66.4\,\mu\mathrm{W}$) already satisfy the target specifications. This indicates a quantitative unmet-specification failure rather than a structural error: the load gate is controlled by a node coupled to \texttt{Vout} through \texttt{R1}, which limits the effective output impedance and caps the achievable gain.

Given the simulation feedback, the model makes a targeted load-network revision without changing the circuit interface or device count. It reconnects \texttt{M1} as a direct PMOS load driven by the external bias \texttt{Vb1}, and converts \texttt{M2} into a diode-connected PMOS device that stabilizes the internal node \texttt{net1}:
\begin{lstlisting}[basicstyle=\ttfamily\small, breaklines=true, columns=fullflexible, keepspaces=true]
M1 (Vout, Vb1,  VDD,  VDD)
M2 (net1, net1, VDD,  VDD)
M3 (Vout, Vb2,  Vin,  VSS)
R1 (net1, Vout)
\end{lstlisting}
After this edit, \texttt{M1} directly drives \texttt{Vout} under the clean external bias \texttt{Vb1}, while \texttt{M2} provides a stable diode-connected bias node rather than acting as the output load controlled by a feedback-coupled internal node. After BO-based sizing on the revised topology, the gain increases from $22.9\,\mathrm{dB}$ to $26.0\,\mathrm{dB}$, while bandwidth ($1.39\,\mathrm{MHz}$) and power ($122\,\mu\mathrm{W}$) still satisfy the target specification. This example shows that \method{} can use simulation feedback to diagnose a performance-limiting load arrangement and repair it through a minimal role reassignment among existing devices.

\subsection{Failure Mode Distribution}
\label{app:failure}
We categorize each evaluation rollout into five mutually exclusive
outcomes according to the validation pipeline.
A candidate is labeled as \textit{invalid syntax} if its netlist
cannot be parsed. If parsing succeeds but the circuit fails topology
or connectivity checks, it is labeled as a \textit{topology failure}.
A structurally valid circuit is labeled as a \textit{BO sizing failure}
if optimization or SPICE evaluation fails to produce valid performance
measurements. If valid simulation results are obtained but at least
one target specification remains unsatisfied, the outcome is
\textit{specification not met}; otherwise, it is labeled as
\textit{success}. Under the two-pass setting, a rollout is counted
as successful if either attempt succeeds; when both attempts fail,
the failure category of the final attempted design is reported.

\begin{table}[h]
\centering
\caption{\textbf{Failure mode distribution (\%) across
different models.}
Generated circuits are categorized into five mutually
exclusive outcomes based on deterministic circuit
validation. Results are reported for single-pass
generation and two-pass generation with simulation
feedback.}
\label{tab:failure_modes}
\resizebox{0.9\linewidth}{!}{
\begin{tabular}{llccccc}
\toprule
\textbf{Setting} & \textbf{Model}
& \textbf{\makecell{Invalid\\Syntax}}
& \textbf{\makecell{Topology\\Failure}}
& \textbf{\makecell{BO Sizing\\Failure}}
& \textbf{\makecell{Spec.\\Not Met}}
& \textbf{Success} \\
\midrule
\multirow{6}{*}{Single-pass}
& Claude-Opus-4.7 & 0.1 & 6.5 & 0.1 & 37.8 & 55.5 \\
& GPT-5.4 & 0.0 & 7.9 & 0.0 & 40.3 & 51.8 \\
& QwQ-32B & 25.7 & 4.6 & 0.0 & 31.1 & 38.6 \\
& Qwen3-32B & 22.3 & 3.6 & 0.0 & 34.1 & 40.0 \\
& gpt-oss-20b & 32.3 & 4.4 & 0.0 & 31.6 & 31.7 \\
& \method{} (Ours) & 0.0 & 4.2
& 1.1 & 23.3 & 71.4 \\
\midrule
\multirow{6}{*}{Two-pass}
& Claude-Opus-4.7 & 0.1 & 3.9 & 0.6 & 25.1 & 70.4 \\
& GPT-5.4 & 1.8 & 0.1 & 0.0 & 31.2 & 66.9 \\
& QwQ-32B & 26.4 & 10.9 & 2.0 & 16.6 & 44.1 \\
& Qwen3-32B & 16.9 & 11.5 & 0.0 & 18.3 & 53.3 \\
& gpt-oss-20b & 36.6 & 5.8 & 0.3 & 16.2 & 41.1 \\
& \method{} (Ours) & 0.0 & 3.8
& 0.2 & 17.8 & 78.2 \\
\bottomrule
\end{tabular}}
\end{table}

\subsection{Topology Diversity and Correction Behavior}

\begin{table}[h]
\centering
\caption{\textbf{Topology diversity and correction behavior.}
We report the number of unique successful circuit topologies generated across five circuit types, where a higher count indicates greater observed diversity under the same evaluation budget. Average device counts before and after correction characterize circuit size. Bold indicates the largest number of unique successful topologies.}
\label{tab:topology_analysis}
\resizebox{0.7\linewidth}{!}{
\begin{tabular}{lccc}
\toprule
\textbf{Model}
& \textbf{\makecell{Unique Successful\\Topologies}}
& \textbf{\makecell{Device Count\\Single-pass}}
& \textbf{\makecell{Device Count\\Two-pass}} \\
\midrule
Claude-Opus-4.7 & 32 & 3.59 & 4.09 \\
GPT-5.4 & 59 & 4.70 & 5.06 \\
\method{} (Ours) & \textbf{81} & 3.54 & 3.50 \\
\bottomrule
\end{tabular}}
\end{table}

%% file: neurips_2026.bib
@article{lai2026analogcoder,
  title={Analogcoder-pro: Unifying analog circuit generation and optimization via multi-modal llms},
  author={Lai, Yao and Poddar, Souradip and Lee, Sungyoung and Chen, Guojin and Hu, Mengkang and Yu, Bei and Luo, Ping and Pan, David Z},
  journal={IEEE Transactions on Computer-Aided Design of Integrated Circuits and Systems},
  year={2026},
  publisher={IEEE}
}

@article{bhandari2024masala,
  title={Masala-chai: A large-scale spice netlist dataset for analog circuits by harnessing ai},
  author={Bhandari, Jitendra and Bhat, Vineet and He, Yuheng and Rahmani, Hamed and Garg, Siddharth and Karri, Ramesh},
  journal={arXiv preprint arXiv:2411.14299},
  year={2024}
}

@article{shao2024deepseekmath,
  title={Deepseekmath: Pushing the limits of mathematical reasoning in open language models},
  author={Shao, Zhihong and Wang, Peiyi and Zhu, Qihao and Xu, Runxin and Song, Junxiao and Bi, Xiao and Zhang, Haowei and Zhang, Mingchuan and Li, YK and Wu, Yang and others},
  journal={arXiv preprint arXiv:2402.03300},
  year={2024}
}

@inproceedings{lai2025analogcoder,
  title={Analogcoder: Analog circuit design via training-free code generation},
  author={Lai, Yao and Lee, Sungyoung and Chen, Guojin and Poddar, Souradip and Hu, Mengkang and Pan, David Z and Luo, Ping},
  booktitle={Proceedings of the AAAI Conference on Artificial Intelligence},
  volume={39},
  number={1},
  pages={379--387},
  year={2025}
}

@article{wang2025analogsage,
  title={AnalogSAGE: Self-evolving Analog Design Multi-Agents with Stratified Memory and Grounded Experience},
  author={Wang, Zining and Gao, Jian and Fu, Weimin and Guo, Xiaolong and Zhang, Xuan},
  journal={arXiv preprint arXiv:2512.22435},
  year={2025}
}

@article{gao2025analoggenie,
  title={Analoggenie: A generative engine for automatic discovery of analog circuit topologies},
  author={Gao, Jian and Cao, Weidong and Yang, Junyi and Zhang, Xuan},
  journal={arXiv preprint arXiv:2503.00205},
  year={2025}
}

@article{chang2024lamagic,
  title={Lamagic: Language-model-based topology generation for analog integrated circuits},
  author={Chang, Chen-Chia and Shen, Yikang and Fan, Shaoze and Li, Jing and Zhang, Shun and Cao, Ningyuan and Chen, Yiran and Zhang, Xin},
  journal={arXiv preprint arXiv:2407.18269},
  year={2024}
}

@article{chang2025lamagic2,
  title={Lamagic2: Advanced circuit formulations for language model-based analog topology generation},
  author={Chang, Chen-Chia and Lin, Wan-Hsuan and Shen, Yikang and Chen, Yiran and Zhang, Xin},
  journal={arXiv preprint arXiv:2506.10235},
  year={2025}
}

@inproceedings{chen2024artisan,
  title={Artisan: Automated operational amplifier design via domain-specific large language model},
  author={Chen, Zihao and Huang, Jiangli and Liu, Yiting and Yang, Fan and Shang, Li and Zhou, Dian and Zeng, Xuan},
  booktitle={Proceedings of the 61st ACM/IEEE Design Automation Conference},
  pages={1--6},
  year={2024}
}

@article{vijayaraghavan2025autocircuit,
  title={Autocircuit-rl: Reinforcement learning-driven llm for automated circuit topology generation},
  author={Vijayaraghavan, Prashanth and Shi, Luyao and Degan, Ehsan and Mukherjee, Vandana and Zhang, Xin},
  journal={arXiv preprint arXiv:2506.03122},
  year={2025}
}

@inproceedings{li2024analoggym,
  title={Analoggym: An open and practical testing suite for analog circuit synthesis},
  author={Li, Jintao and Zhi, Haochang and Lyu, Ruiyu and Li, Wangzhen and Bi, Zhaori and Zhu, Keren and Zeng, Yanhan and Shan, Weiwei and Yan, Changhao and Yang, Fan and others},
  booktitle={Proceedings of the 43rd IEEE/ACM International Conference on Computer-Aided Design},
  pages={1--9},
  year={2024}
}

@article{shi2025amsnet,
  title={AMSnet-KG: A netlist dataset for LLM-based AMS circuit auto-design using knowledge graph RAG},
  author={Shi, Yichen and Tao, Zhuofu and Gao, Yuhao and Zhou, Tianjia and Chang, Cheng and Wang, Yaxin and Chen, Bingyu and Zhang, Genhao and Liu, Alvin and Yu, Zhiping and others},
  journal={ACM Transactions on Design Automation of Electronic Systems},
  volume={30},
  number={6},
  pages={1--37},
  year={2025},
  publisher={ACM New York, NY}
}

@article{dutta2026can,
  title={Can an Actor-Critic Optimization Framework Improve Analog Design Optimization?},
  author={Dutta, Sounak and Amin, Fin and Panda, Sushil and Rabe, Jonathan and Wen, Yuejiang and Franzon, Paul},
  journal={arXiv preprint arXiv:2603.24714},
  year={2026}
}

@inproceedings{wang2020gcn,
  title={GCN-RL circuit designer: Transferable transistor sizing with graph neural networks and reinforcement learning},
  author={Wang, Hanrui and Wang, Kuan and Yang, Jiacheng and Shen, Linxiao and Sun, Nan and Lee, Hae-Seung and Han, Song},
  booktitle={2020 57th ACM/IEEE Design Automation Conference (DAC)},
  pages={1--6},
  year={2020},
  organization={IEEE}
}

@inproceedings{poddar2025insight,
  title={INSIGHT: A Universal Neural Simulator Framework for Analog Circuits with Autoregressive Transformers},
  author={Poddar, Souradip and Oh, Youngmin and Lai, Yao and Zhu, Hanqing and Hwang, Bosun and Pan, David Z},
  booktitle={2025 62nd ACM/IEEE Design Automation Conference (DAC)},
  pages={1--7},
  year={2025},
  organization={IEEE}
}

@inproceedings{zhang2025analogxpert,
  title={Analogxpert: Automating analog topology synthesis by incorporating circuit design expertise into large language models},
  author={Zhang, Haoyi and Sun, Shizhao and Lin, Yibo and Wang, Runsheng and Bian, Jiang},
  booktitle={2025 International Symposium of Electronics Design Automation (ISEDA)},
  pages={772--777},
  year={2025},
  organization={IEEE}
}

@article{mehradfar2024aicircuit,
  title={Aicircuit: A multi-level dataset and benchmark for ai-driven analog integrated circuit design},
  author={Mehradfar, Asal and Zhao, Xuzhe and Niu, Yue and Babakniya, Sara and Alesheikh, Mahdi and Aghasi, Hamidreza and Avestimehr, Salman},
  journal={arXiv preprint arXiv:2407.18272},
  year={2024}
}

@article{dong2023cktgnn,
  title={Cktgnn: Circuit graph neural network for electronic design automation},
  author={Dong, Zehao and Cao, Weidong and Zhang, Muhan and Tao, Dacheng and Chen, Yixin and Zhang, Xuan},
  journal={arXiv preprint arXiv:2308.16406},
  year={2023}
}

@article{nagel1973spice,
  title={SPICE (simulation program with integrated circuit emphasis)},
  author={Nagel, Laurence and Pederson, Donald O},
  year={1973}
}

@article{mcconaghy2011trustworthy,
  title={Trustworthy genetic programming-based synthesis of analog circuit topologies using hierarchical domain-specific building blocks},
  author={McConaghy, Trent and Palmers, Pieter and Steyaert, Michiel and Gielen, Georges GE},
  journal={IEEE Transactions on Evolutionary Computation},
  volume={15},
  number={4},
  pages={557--570},
  year={2011},
  publisher={IEEE}
}

@article{zhao2020automated,
  title={An automated topology synthesis framework for analog integrated circuits},
  author={Zhao, Zhenxin and Zhang, Lihong},
  journal={IEEE Transactions on Computer-Aided Design of Integrated Circuits and Systems},
  volume={39},
  number={12},
  pages={4325--4337},
  year={2020},
  publisher={IEEE}
}

@article{abel2022fuboco,
  title={FUBOCO: Structure synthesis of basic op-amps by FUnctional BlOck COmposition},
  author={Abel, Inga and Graeb, Helmut},
  journal={ACM Transactions on Design Automation of Electronic Systems (TODAES)},
  volume={27},
  number={6},
  pages={1--27},
  year={2022},
  publisher={ACM New York, NY}
}

@article{zhao2024automated,
  title={Automated topology synthesis of analog integrated circuits with frequency compensation},
  author={Zhao, Zhenxin and Liu, Jun and Zhao, Wen-Sheng and Zhang, Lihong},
  journal={IEEE Transactions on Computer-Aided Design of Integrated Circuits and Systems},
  volume={44},
  number={3},
  pages={832--844},
  year={2024},
  publisher={IEEE}
}

@article{zhao2022analog,
  title={Analog integrated circuit topology synthesis with deep reinforcement learning},
  author={Zhao, Zhenxin and Zhang, Lihong},
  journal={IEEE Transactions on Computer-Aided Design of Integrated Circuits and Systems},
  volume={41},
  number={12},
  pages={5138--5151},
  year={2022},
  publisher={IEEE}
}

@misc{qwq32b,
    title = {QwQ-32B: Embracing the Power of Reinforcement Learning},
    url = {https://qwenlm.github.io/blog/qwq-32b/},
    author = {Qwen Team},
    month = {March},
    year = {2025}
}

@misc{qwen3technicalreport,
      title={Qwen3 Technical Report}, 
      author={Qwen Team},
      year={2025},
      eprint={2505.09388},
      archivePrefix={arXiv},
      primaryClass={cs.CL},
      url={https://arxiv.org/abs/2505.09388}, 
}

@misc{openai2025gptoss120bgptoss20bmodel,
      title={gpt-oss-120b \& gpt-oss-20b Model Card}, 
      author={OpenAI},
      year={2025},
      eprint={2508.10925},
      archivePrefix={arXiv},
      primaryClass={cs.CL},
      url={https://arxiv.org/abs/2508.10925}, 
}

@misc{openai2025gpt52,
  title        = {Introducing GPT-5.2},
  author       = {{OpenAI}},
  year         = {2025},
  howpublished = {\url{https://openai.com/index/introducing-gpt-5-2/}},
  note         = {Accessed: 2026-05-05}
}

@misc{openai2026gpt54,
  title        = {Introducing GPT-5.4},
  author       = {{OpenAI}},
  year         = {2026},
  howpublished = {\url{https://openai.com/index/introducing-gpt-5-4/}},
  note         = {Accessed: 2026-05-05}
}

@misc{anthropic2026opus47,
  title        = {Introducing Claude Opus 4.7},
  author       = {{Anthropic}},
  year         = {2026},
  howpublished = {\url{https://www.anthropic.com/news/claude-opus-4-7}},
  note         = {Accessed: 2026-05-05}
}

@inproceedings{blocklove2023chip,
  title={Chip-chat: Challenges and opportunities in conversational hardware design},
  author={Blocklove, Jason and Garg, Siddharth and Karri, Ramesh and Pearce, Hammond},
  booktitle={2023 ACM/IEEE 5th Workshop on Machine Learning for CAD (MLCAD)},
  pages={1--6},
  year={2023},
  organization={IEEE}
}

@article{chang2023chipgpt,
  title={Chipgpt: How far are we from natural language hardware design},
  author={Chang, Kaiyan and Wang, Ying and Ren, Haimeng and Wang, Mengdi and Liang, Shengwen and Han, Yinhe and Li, Huawei and Li, Xiaowei},
  journal={arXiv preprint arXiv:2305.14019},
  year={2023}
}

@article{liu2023chipnemo,
  title={Chipnemo: Domain-adapted llms for chip design},
  author={Liu, Mingjie and Ene, Teodor-Dumitru and Kirby, Robert and Cheng, Chris and Pinckney, Nathaniel and Liang, Rongjian and Alben, Jonah and Anand, Himyanshu and Banerjee, Sanmitra and Bayraktaroglu, Ismet and others},
  journal={arXiv preprint arXiv:2311.00176},
  year={2023}
}

@inproceedings{tang2026towards,
  title={Towards high data efficiency in reinforcement learning with verifiable reward},
  author={Tang, Xinyu and Zhang, Zhenduo and Liu, Yurou and Zhao, Xin and Zhang, Zhiqiang and ZHOU, JUN and others},
  booktitle={International Conference on Learning Representations},
  volume={2026},
  pages={32611--32635},
  year={2026}
}

@inproceedings{nguyen2026adaptive,
  title={Adaptive rollout allocation for online reinforcement learning with verifiable rewards},
  author={Nguyen, Trung Hieu and Nguyen, Bao and Ma, Wenao and Zhao, Yuzhi and She, Ruifeng and Nguyen, Viet Anh},
  booktitle={International Conference on Learning Representations},
  volume={2026},
  pages={136100--136132},
  year={2026}
}

@article{sun2026improving,
  title={Improving data efficiency for llm reinforcement fine-tuning through difficulty-targeted online data selection and rollout replay},
  author={Sun, Yifan and Shen, Jingyan and Wang, Yibin and Chen, Tianyu and Wang, Zhendong and Zhou, Mingyuan and Zhang, Huan},
  journal={Advances in Neural Information Processing Systems},
  volume={38},
  pages={173449--173476},
  year={2026}
}

@article{grattafiori2024llama,
  title={The llama 3 herd of models},
  author={Grattafiori, Aaron and Dubey, Abhimanyu and Jauhri, Abhinav and Pandey, Abhinav and Kadian, Abhishek and Al-Dahle, Ahmad and Letman, Aiesha and Mathur, Akhil and Schelten, Alan and Vaughan, Alex and others},
  journal={arXiv preprint arXiv:2407.21783},
  year={2024}
}

@article{glm2024chatglm,
  title={Chatglm: A family of large language models from glm-130b to glm-4 all tools},
  author={Glm, Team and Zeng, Aohan and Xu, Bin and Wang, Bowen and Zhang, Chenhui and Yin, Da and Zhang, Dan and Rojas, Diego and Feng, Guanyu and Zhao, Hanlin and others},
  journal={arXiv preprint arXiv:2406.12793},
  year={2024}
}
